\documentclass[]{pasj02} 
\usepackage{url}

\jyear{2025}
\Received{}
\Accepted{}

\begin{document}

\title{XRISM analysis of the complex Fe K$\alpha$ line in Centaurus A}

\author{
 David \textsc{Bogensberger}\altaffilmark{1,2}\altemailmark\orcid{0000-0002-5924-4822} \email{david.bogensberger2@gmail.com}, 
 Yuya  \textsc{Nakatani}\altaffilmark{3}\orcid{0009-0000-9577-8701},
 Tahir \textsc{Yaqoob}\altaffilmark{4,5,6},
 Yoshihiro \textsc{Ueda}\altaffilmark{3}\orcid{0000-0001-7821-6715},
 Richard \textsc{Mushotzky}\altaffilmark{7,8}\orcid{0000-0002-7962-5446},
 Jon M. \textsc{Miller}\altaffilmark{2}\orcid{0000-0003-2869-7682},
 Luigi C. \textsc{Gallo}\altaffilmark{9}\orcid{0009-0006-4968-7108},
 Yasushi \textsc{Fukazawa}\altaffilmark{10}\orcid{0000-0002-0921-8837},
 Taishu \textsc{Kayanoki}\altaffilmark{10}\orcid{0000-0002-6960-9274},
 Makoto \textsc{Tashiro}\altaffilmark{11,12}\orcid{0000-0002-5097-1257},
 Hirofumi \textsc{Noda}\altaffilmark{13}\orcid{0000-0001-6020-517X},
 Toshiya \textsc{Iwata}\altaffilmark{14}\orcid{0000-0002-2304-4773},
 Kouichi \textsc{Hagino}\altaffilmark{14}\orcid{0000-0003-4235-5304},
 Misaki \textsc{Mizumoto}\altaffilmark{15}\orcid{0000-0003-2161-0361},
 Misaki \textsc{Urata}\altaffilmark{8},
 Frederick S. \textsc{Porter}\altaffilmark{4}\orcid{0000-0002-6374-1119},
 and 
 Michael \textsc{Loewenstein}\altaffilmark{7,4,5}\orcid{0000-0002-1661-4029}
}
\altaffiltext{1}{Universit\'e Paris-Saclay, Universit\'e Paris Cit\'e, CEA, CNRS, AIM, 91191 Gif-sur-Yvette, France}
\altaffiltext{2}{Department of Astronomy, The University of Michigan, 1085 South University Avenue, Ann Arbor, Michigan, 48103, USA}
\altaffiltext{3}{Department of Astronomy, Kyoto University, Kitashirakawa-Oiwake-cho, Sakyo-ku, Kyoto 606-8502, Japan}
\altaffiltext{4}{National Aeronautics and Space Administration, Goddard Space Flight Center, Greenbelt, MD 20771, USA}
\altaffiltext{5}{Center for Research and Exploration in Space Science and Technology, NASA/GSFC (CRESST II), Greenbelt, MD 20771, USA}
\altaffiltext{6}{Center for Space Science and Technology, University of Maryland, Baltimore County (UMBC), 1000 Hilltop Circle, Baltimore, MD 21250, USA}
\altaffiltext{7}{Department of Astronomy, University of Maryland, College Park, MD 20742, USA}
\altaffiltext{8}{Joint Space-Science Institute, University of Maryland, College Park, MD 20742, USA}
\altaffiltext{9}{Saint Mary’s University, Department of Astronomy and Physics, 923 Robie St Halifax, Nova Scotia B3H 3C3, Canada}
\altaffiltext{10}{Department of Physics, Graduate School of Advanced Science and Engineering, Hiroshima University, Hiroshima 739-8526, Japan}
\altaffiltext{11}{Department of Physics, Saitama University, Saitama 338-8570, Japan}
\altaffiltext{12}{Institute of Space and Astronautical Science (ISAS), Japan Aerospace Exploration Agency (JAXA), Kanagawa 252-5210, Japan}
\altaffiltext{13}{Astronomical Institute, Tohoku University, 6-3 Aramakiazaaoba, Aoba-ku, Sendai, Miyagi 980-8578, Japan}
\altaffiltext{14}{Department of Physics, The University of Tokyo, 7-3-1 Hongo, Bunkyo-ku, Tokyo 113-0033, Japan}
\altaffiltext{15}{Science Education Research Unit, University of Teacher Education Fukuoka, Akama-bunkyo-machi, Munakata, Fukuoka 811-4192, Japan}



\maketitle

\begin{abstract}

We analyze the high-resolution XRISM/Resolve spectrum of the Fe K$\alpha$ emission line of the nearest active galactic nucleus, in Centaurus A. The line features two narrow and resolved peaks of Fe K$\alpha_1$, and Fe K$\alpha_2$ with a FWHM of $(4.8\pm0.2)\times10^2~\mathrm{km}~\mathrm{s}^{-1}$ each. A broad line with a FWHM of $(4.3\pm0.3)\times10^3~\mathrm{km}~\mathrm{s}^{-1}$, and with a flux similar to the two narrow line cores, is also required. This broad component is not observed in the optical or IR spectra of Cen A. The line shape requires the existence of an emission region that extends from $\sim10^{-3}~\mathrm{pc}$ to $\sim10^1~\mathrm{pc}$. Assuming that the emissivity follows a radial power-law profile of $r^{-q}$, we find $q\approx2$. This may indicate an extended corona, an emitting region that bends towards the corona, or a non-uniform density. When assuming $q=3$, the line shape can only be reproduced by including three emitting components in the model. The measured best-fit inclination is $ 24^{+13}_{-7}~\degree$, but higher inclinations are only slightly disfavored. A single blurred \texttt{MYTorusL} line profile can describe the line shape, but requires a large relative normalization. This could be due to past variability, modified abundances, or differing geometries. The line shape can be reproduced from the radii measured by reverberation mapping, but only if an additional extended emitting region at small radii is included.

\end{abstract}


\section{Introduction}\label{sec:intro}

The nearest of all the giant radio galaxies, Centaurus A (NGC 5128, henceforth Cen A), provides a unique opportunity to observe the dynamics of an active galaxy as traced by the Fe K$\alpha$ complex in detail. As one of the brightest active galactic nuclei (AGNs) in the X-ray band \citep{2009A&A...505..417B, 2011PASJ...63S.677H} it has been extensively studied by virtually every X-ray observatory. The Resolve microcalorimeter \citep{Porter24} on XRISM \citep{Tashiro20, Tashiro24} provides the capability to achieve a full width at half maximum energy resolution (FWHM) of $\mathrm{FWHM} < 5 ~\mathrm{eV}$ at $\sim7~\mathrm{keV}$, which enables a detailed examination of the Fe K$\alpha$ complex.

The interaction of the nuclear continuum radiation with distinct parts of the accretion flow and surrounding medium can produce distinct emission and absorption lines in AGNs. The measured energy and exact shape of these lines can trace the composition, geometry, and kinematics of the central regions. For instance, the Doppler effect can broaden lines originating from material closer to the black hole more than from material originating further away because of greater orbital speeds.  For lines originating in material very close to the black hole, special and general relativistic effects can further distort the observed line profiles. 

The most prominent of the emission lines visible in the X-ray spectra of AGNs is the Fe K$\alpha$ fluorescence line arising from a 2p-1s electron transition. The spin-orbit interaction creates two components; Fe K$\alpha_1$ at $6.404~\mathrm{keV}$, and Fe K$\alpha_2$ at $6.391~\mathrm{keV}$ \citep{1997PhRvA..56.4554H} with an intensity ratio of 2:1. Previous instruments lacked the resolving power to distinguish these two lines, and it was normally sufficient to model the features with a single-line profile at an energy of $\sim6.4~\mathrm{keV}$. However, XRISM/Resolve has sufficient sensitivity and spectral resolution to resolve both features, as well as other narrow spectral features \citep{2023arXiv230210930G, 2024ApJ...973L..25X}.

The Fe K$\beta$ line at $7.058~\mathrm{keV}$ is also detected in many X-ray spectra. It is significantly weaker than the Fe K$\alpha$ line and it is found just below the neutral Fe K-edge at $7.112~\mathrm{keV}$. \citet{2024ApJ...961..150B} also reported the detection of Fe XXV and Fe XXVI absorption lines in the Chandra spectra of Cen A, and determined the velocity shifts of the lines to vary over time. In different AGNs, the He-like and H-like lines of Fe can be observed in emission \citep{2002A&A...387...76B} or absorption \citep{2005MNRAS.357..599B}, depending on the location of the highly ionized material relative to the line of sight of the observer.

The dependence of the Fe K$\alpha$ emissivity on the radius at which it is emitted is typically treated as a power law with $\epsilon (r) \propto r^{-q}$. Measurements of the index $q$ can constrain the location of the emitting material relative to the ionizing source, the corona. Theoretical emissivity profiles of accretion disks find that $q>3$ very close to the black hole, if $r\lesssim 30~r_{\rm g}$ \citep{2012MNRAS.424.1284W, 2012A&A...545A.106S, 2017MNRAS.472.1932G}. At larger radii, $q$ is expected to approach $\approx3$ \citep{2024MNRAS.532.3786Z}. It has also been argued that it could eventually drop to $q\approx2$ \citep{2016ApJ...826...52K}. A value of $q=3$ is the result of the inverse-square dependence of flux on radius, and an extra factor of the radius to account for the angle between the corona and the disk, at large radii.

\citet{2024ApJ...973L..25X} previously investigated two XRISM/Resolve spectra of the AGN NGC 4151, and found that its Fe K$\alpha$ line could be described by three distinct components at different radii from the black hole. Using the \texttt{MYTorusL} model to describe the line shapes, it was found that these three components also had slightly different inclinations and Hydrogen column densities, corresponding to distinct elements of the accretion flow. They associated these three components with the disk, the broad-line region, and the torus. This result raises the question of whether other AGNs have similar line profiles that require multiple independent emission regions. 
 
Cen A is the nearest AGN at $3.8\pm0.1~\mathrm{Mpc}$ \citep{2010PASA...27..457H}. It is a LINER/Seyfert-2 and is also a FR I radio galaxy \citep{2010ApJ...708..675G}, whose jet is prominently seen in X-rays \citep{2003ApJ...593..169H, 2024ApJ...974..307B}. There has been significant debate about the inclination of the jet and the accretion flow because of contradictory measurements resulting from different methods and datasets. Its classification as a Type-2 AGN or LINER suggests a high inclination. However, the difference in the brightness of the jet and counterjet suggests a low inclination \citep{2003ApJ...593..169H}. Recent measurements using very-long-baseline interferometry \citep{2021NatAs...5.1017J, 2025AJ....169...37P} on mpc scales, and proper motions on kpc scales \citep{2024ApJ...974..307B} have preferred lower inclinations of $12-45\degree$, $<25\degree$, and $<41\degree$, respectively. A low jet inclination was also measured by \citet{2003ApJ...593..169H} and \citet{2014A&A...569A.115M}. In contrast, some previous investigations found high jet inclinations between $50\degree$ and $80\degree$ \citep{1994ApJ...426L..23S, 1996ApJ...466L..63J, 1998AJ....115..960T}.

The outer molecular disk at radii of tens of parsecs appears to possess a high inclination of $70\degree$ \citep{2009ApJ...695..116E}, and even $>85\degree$ \citep{2009ApJ...702.1127R}. However, the inclination in more central regions, at radii of a few parsecs, is found to exhibit lower inclination values of $25-40\degree$ \citet{2010PASA...27..396Q}, $25\degree$ \citep{2007MNRAS.374..385K}, and $34\degree$ \citep{2007ApJ...671.1329N}.  This indicates that the outer disk is likely warped. At small radii, the accretion flow appears to have approximately the same inclination as the jet. At large radii, the outer molecular disk aligns more with the inclination of the galaxy, of $72\pm3\degree$ \citep{1979AJ.....84..284D}. 

Warps have been theoretically studied by \citet{1999MNRAS.304..557O, 2014MNRAS.441.1408T, 2023NewA..10102012L, 2023ApJ...944L..48L}. The inclination of the Fe K$\alpha$ line-emitting region could not be investigated yet because of the limited spectral resolution of previous X-ray missions. It might not even be directly associated with accretion flow and might originate from a wind \citep{1994MNRAS.266..653Z} or large-scale structures such as the optical/IR absorption lane in Cen A. Previous investigations of the X-ray spectra of Cen A lacked the sensitivity to distinguish between the wide range of possible interpretations.

Past observations have shown that the Fe K$\alpha$ line in Cen A is relatively narrow and has one of the highest fluxes for an AGN. This makes it an ideal target for investigating the origin of this emission.   The geometry of the accretion flow or other possible origins can be ascertained with careful measurement of the line profile. However, the complexity of the line profile has long been recognized.  Simultaneous fitting to BeppoSAX and COMPTEL spectra showed the Fe band is best described by combining a narrow Fe K$\alpha$ line and a broad component at $6.8~\mathrm{keV}$ \citep{2003ApJ...593..160G}. However, this shape was not observed in subsequent observations. \citet{2024ApJ...961..150B} also found the Fe K$\alpha$ line to be slightly redshifted compared to the systemic velocity of the host galaxy. 

Several teams have independently and consistently measured the redshift of Cen A galaxy using different methods and different datasets \citep{1978PASP...90..237G, 1983BAAS...15..921W, 1995ApJ...449..592H, 2006AJ....131.1163S, 2007AJ....134..494W, 2015A&A...574A.109W}. The average of these measurements is $(1.819\pm0.010)\times10^{-3}$ or $545~\mathrm{km}~\mathrm{s}^{-1}$.

This paper focuses on an analysis of the Fe K$\alpha$ line from the first  XRISM/Resolve X-ray spectrum of Cen A, observed on 2024 August 4. Future papers will investigate other features of this dataset and analyze the entire energy range together with NuSTAR and XRISM/Xtend spectra.

This paper is structured as follows. Section \ref{sec:obs} describes the observations used in this work and how the data were reduced and analyzed. In Section \ref{sec:Feline}, we investigate the properties of the Fe K$\alpha$ line in the $6.0-6.8~\mathrm{keV}$ band as observed by XRISM/Resolve, and describe several different approaches to fitting it. This is followed by Section \ref{sec:caveats}, in which we consider some caveats of these models and discuss possible future steps to improve and refine the analysis. In Section \ref{sec:disc}, we discuss our main results and compare them to previous investigations. Finally, we present our conclusions in Section \ref{sec:conc}.

\section{Observations and Data Analysis}\label{sec:obs}

\subsection{XRISM}\label{sec:XRISMobs}

Cen A was observed by XRISM on 2024 August 4, for a duration of $\sim 4.82$~days (observation ID 300019010) (Table~\ref{tab:first}). The data from the XRISM pipeline were reprocessed with XRISM software in the HEASoft release version 6.34 and the XRISM CalDB 9 version of the XRISM calibration database. Post-pipeline analysis was also performed with these versions of the XRISM software and CalDB. The reprocessing was performed on data produced by the JAXA pre-pipeline v03.01. The cleaned Resolve event file consisted of a net exposure time of $227.99~\mathrm{ks}$, beginning at UT 19:17:04, and had a mean net $0-30~\mathrm{keV}$ count rate of $\approx1.98 ~ \mathrm{cts} ~ \mathrm{s}^{-1}$. 

In XRISM CalDB 9, the Resolve line-spread-function (LSF) calibration file is based on fine-tuning using in-flight data. The cleaned Resolve event file has standard screening applied by the pipeline, corresponding to the label ``PIXELALL'' in the screening CalDB file. Post-pipeline screening and extraction of data products were performed following the Quick-Start Guide, v2.1 (hereafter, QSG \footnote{\url{https://heasarc.gsfc.nasa.gov/docs/xrism/analysis/quickstart/index.html}}), except that ``proximity screening'' (by means of the ``STATUS[4]'' flag) was not applied. Proximity screening aims to remove frame events caused by cosmic rays by flagging events with times that are closer in arrival time to each other than expected from the Poisson probability, given the source count rate. However, there is an increasing fraction of false positives with increasing source count rate, so real source events are removed uniformly over the entire energy band. Since frame events occur predominantly below $\sim 2~\mathrm{keV}$, the proximity screen was not applied, albeit resulting in saving only $\sim 0.3\%$ of the source photons. 

Two spectral response matrices (RMFs) were constructed, one small (``S'') and one large (``L''). The former consists only of the core Gaussian LSF, whilst the latter includes the exponential tail, the escape peaks, and the Si fluorescence line (both exclude the ``ELC'', or electron-loss continuum). To mitigate the distortion of the effective area due to false low-resolution secondary events (Ls), the RMFs were made by only inputting non-Ls $3-10~\mathrm{keV}$ events into the RMF generator\footnote{\url{https://heasarc.gsfc.nasa.gov/docs/xrism/analysis/abc_guide/xrism_abc.pdf}}. If some of the Ls events are real, the systematic effect on the effective area is that it could be lower by $\sim 2\%$ (at all energies). 

The ancillary response file (ARF) was made for a single attitude bin, as described in the QSG. In the paper, we utilize data only in the  $5.8-7.6~\mathrm{keV}$ energy band. We examined the ratios of spectra in the ``inner ring'' and ``outer ring'' pixels to the central four pixels to check for sub-array spectral distortion that has been observed in some sources. The inner ring consists of pixels that have one edge adjacent to one of the four central pixels (those with ID 0, 17, 18, or 35), and the outer ring consists of pixels that have one edge adjacent to an inner ring pixel. We found that the spectra from the three groups of pixels are consistent with each other in the $5.8-7.6~\mathrm{keV}$ band. The spectra have not been rebinned.

\begin{table}
\renewcommand{\arraystretch}{1.2}
  \tbl{Properties of the observations used in this paper}{
  \begin{tabular}{cccc}
      \hline
      Telescope & ObsID & Start date & Exposure time (ks)  \\ \hline
      XRISM & 300019010 & 2024-08-04 19:17:04 & 227.99 \\
      Chandra & 29490 & 2024-08-07 06:50:43 & 20.38 \\ \hline
    \end{tabular}}\label{tab:first}
\end{table}

\subsection{Chandra}\label{sec:Chandraobs}

\begin{figure}[t]
\begin{center}
\resizebox{\hsize}{!}{\includegraphics{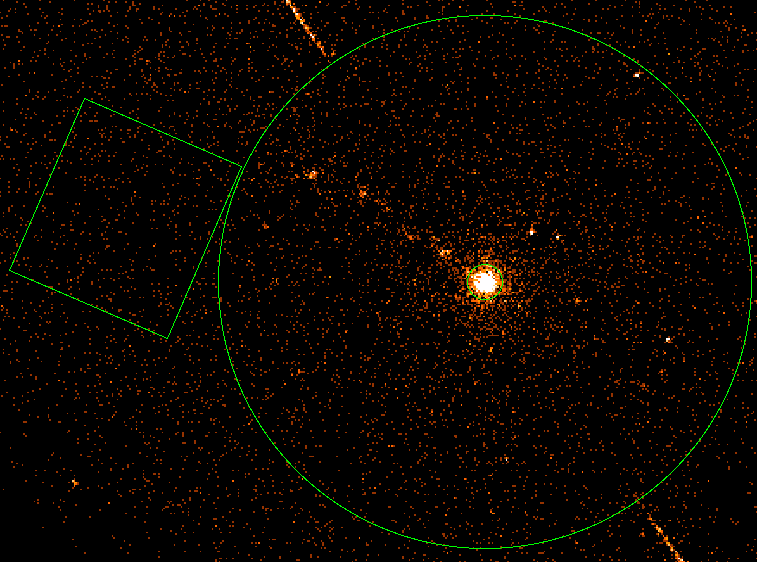}}
\end{center}
\caption{Chandra image of Cen A in observation 29490. The green annulus has inner and outer radii of $5''$ and $76''$, and depicts the source extraction region to determine the spectrum of the contaminating sources in the XRISM/Resolve spectrum. The background extraction region is indicated as a rectangle. Individual pixels are colored according to the number of counts in them, with the color bar covering the interval of 0 to 5 counts per pixel. 
{Alt text: Image of the Chandra observation of Cen A, featuring the bright point-like nucleus, and the jet pointing towards the north-east. Several faint point-like sources are X-ray binaries in Cen A. A green annulus includes the jet and X-ray binaries, but excludes the bright nucleus in the center. The rectangle is drawn further out, and does not include the jet or any X-ray binary.} 
 \label{fig:img}}
\end{figure}

XRISM/Resolve has a half-power diameter of $\approx 1'.3$ \citep{2024arXiv240619911M}. The nucleus is by far the brightest X-ray source of Cen A. However, the X-ray jet, X-ray binaries, supernova remnants, and diffuse galactic X-ray emission from within $1'.3$ of the nucleus also contribute to the observed emission. As we only want to fit the spectrum emitted by the nucleus, we need to account for this contaminating X-ray emission. Chandra has high angular resolution, so we can estimate the properties of the contaminating emission from its observation on 2024 August 7. 

For this data analysis, we used CIAO version 4.16.0. Initially, we removed the readout streak with the task \texttt{acisreadcorr}, with parameters of \texttt{dx}~=~4 and \texttt{dy}~=~50. Next, we extracted a spectrum from an annulus around the nucleus with inner and outer radii of $5''$ and $76''$. These radii were selected to exclude almost all of the nuclear PSF, as well as the grating arms (see Fig. \ref{fig:img}). This annulus includes almost all of the contaminating sources that affect the XRISM/Resolve spectrum. We selected a background corresponding to a region further away from the nucleus, which did not contain the jet or counterjet, the grating arms, or any sources detected by the CIAO task \texttt{wavdetect}. The background is in the part of the image with a similar exposure as the source region. This specific background region is bounded by the green square in Fig. \ref{fig:img}.  

\begin{figure}[t]
\begin{center}
\resizebox{\hsize}{!}{\includegraphics{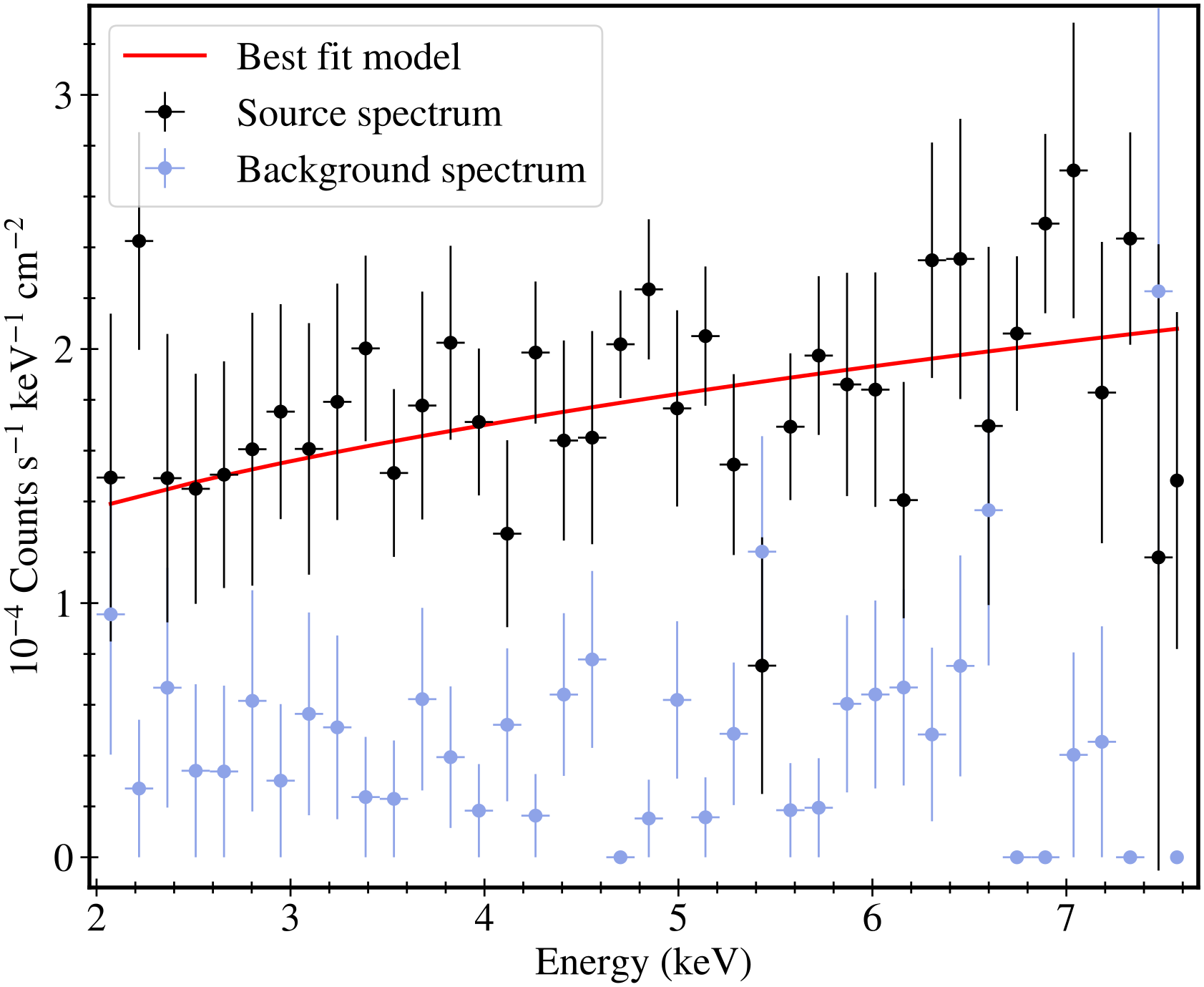}}
\end{center}
\caption{Chandra spectrum of the contaminating X-ray sources which cannot be resolved from the nuclear spectrum observed by XRISM/Resolve. The spectrum is best fitted with a power-law model.  
{Alt text: Plot depicting the background-subtracted source spectrum, as well as the background spectrum of the contaminating X-ray sources. The best-fitting power-law model used to fit the source spectrum is shown as well.} 
 \label{fig:contspec}}
\end{figure}

The X-ray spectrum of the contaminating sources is shown in Fig. \ref{fig:contspec}. We fitted the source spectrum in the energy range $2.0-7.6~\mathrm{keV}$, as the background exceeded the source spectrum at higher and lower energies. 

To estimate the spectral shape of the contaminating emission, we fit the spectrum with a simple power-law model. This simple model was statistically preferred over more complex models based on the calculated Bayesian Information Criterion (BIC; \cite{1978AnSta...6..461S}). The best-fit power-law parameters were found to be a power-law slope of $\Gamma=-0.32\pm0.08$, and a normalization of $N_{c}=\left(1.1\pm0.1\right)\times10^{-4}~\mathrm{cts}~\mathrm{s}^{-1}~\mathrm{keV}^{-1}~\mathrm{cm}^{-2}$ at $1~\mathrm{keV}$. Extrapolating this best-fit, we find that it corresponds to a $2-10 ~\mathrm{keV}$ flux of $(1.53\pm0.06)\times10^{-11} ~\mathrm{erg}~\mathrm{cm}^{-2}~\mathrm{s}^{-1}$. Within the $6.0-6.8~\mathrm{keV}$ band, the contaminating emission has a flux of $(1.62\pm0.02)\times10^{-12}~\mathrm{erg}~\mathrm{cm}^{-2}~\mathrm{s}^{-1}$, which corresponds to $8.4\pm0.3\%$ of the total flux in this band. All subsequent spectral analysis will include this spectrum of the contaminating sources, with the parameters fixed to these values.

\subsection{Spectral fitting}

All spectra were fitted in XSPEC \citep{1996ASPC..101...17A} version 12.14.0h. We assumed solar abundances \citep{2000ApJ...542..914W}, and the photoionization cross-sections of \citet{1996ApJ...465..487V}. The best fits were found by minimizing the C-statistic \citep{1979ApJ...228..939C}. To reduce computing time, we initially fitted the XRISM/Resolve spectrum with the small RMF. Afterwards, we applied the fit to a spectrum linked to the large RMF and fitted the data again, to determine the final best-fit values and their errors. The difference between the best fits with the two different RMFs was negligible. There was no indication that this method affected any of the results. Throughout this paper, we present parameters with  $1\sigma$ uncertainties. 

\section{Fe K$\alpha$ line}\label{sec:Feline}

The core of the Fe K$\alpha$ line has a double peak, corresponding to Fe K$\alpha_1$, and Fe K$\alpha_2$. The intensity drops rapidly on either side of the double peak, but then extends into a broad base. In this section, we fitted the line and the continuum around it between $6.0-6.8 ~\mathrm{keV}$, to understand the geometry of the line-emitting region.

By focusing on this range of energies, there is a risk that the parameters of the Fe K$\alpha$ line are affected by a continuum that tries to fit the line profile as well. The photon index is used in other models that describe the Fe K$\alpha$ line, so an incorrect value of $\Gamma$ can affect our ability to fit the line. This could be alleviated by fitting a larger energy range, but that would necessitate the inclusion of several other spectral parameters, and focus the fits away from an analysis of the Fe K$\alpha$ line. 

To avoid such effects, we first fitted the broader energy range in the Fe band, between $5.8-7.6~\mathrm{keV}$, excluding all bins featuring the Fe K$\alpha$ and Fe K$\beta$ lines. We used an absorbed power-law fit that included five Gaussians, for redshifted and blueshifted Fe XXV and Fe XXVI lines, and the Ni K$\alpha$ line. In this fit, we found a best-fit photon index of $\Gamma=1.95\pm0.03$, and $N_{\rm H} = 34^{+2}_{-1}\times10^{22}~\mathrm{cm}^{-2}$. The best-fit properties of the other emission lines in this energy range will be described in a future paper. Although these best-fit values provide a great description of the continuum in this energy band, they are not consistent with the broad-band X-ray spectrum. The best-fit $N_{\rm H}$ is significantly overestimated, $\Gamma$ is slightly overestimated. These values are, therefore, only used to approximate the continuum in this energy range, not to investigate the physical properties in the system. 

When fitting the $6.0-6.8~\mathrm{keV}$ energy range, we constrained the photon index to be within the $1\sigma$ errors of the best-fit value found from the Fe band continuum. However, we allowed the normalization of the power-law component to vary, to allow for some freedom in the continuum model. We allowed the $N_{\rm H}$ to vary freely, but it was always found to be consistent with the best-fit value of the Fe band continuum. The redshifted and blueshifted Fe XXV lines were also included in the continuum model, and their normalizations were allowed to vary freely. 

\subsection{Gaussian fits}

When fitting the Fe K$\alpha$ line with a single Gaussian function above the continuum, we found that the best fit standard deviation is $\sigma=11.6\pm0.2~\mathrm{eV} = (5.5\pm0.1)\times10^2~\mathrm{km}~\mathrm{s}^{-1}$. However, this model is a poor description of the line profile, and non-physical since the Fe K$\alpha$ line is resolved as a doublet in these spectra. 

A total of three Gaussian profiles were required to accurately describe the line profile. These corresponded to the Fe K$\alpha_1$, Fe K$\alpha_2$ lines, and the broad line. Additional Gaussian profiles only provided minimal improvements in the fit statistic and were ruled out as necessary components of the line profile by the BIC. 

\begin{figure}[t]
\begin{center}
\resizebox{\hsize}{!}{\includegraphics{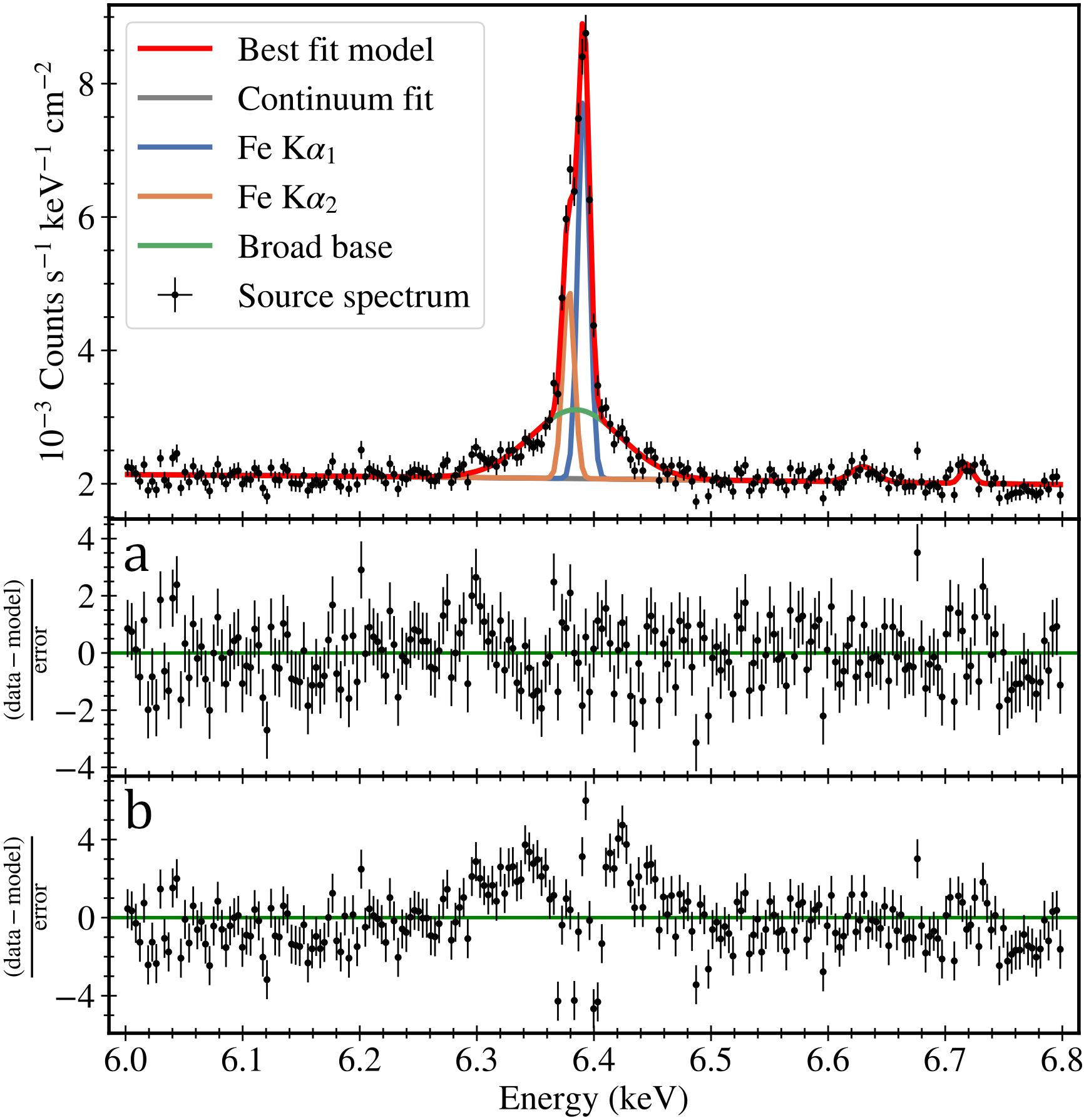}}
\end{center}
\caption{XRISM/Resolve spectrum of the Fe K$\alpha$ line, fitted using three Gaussian profiles, describing the Fe K$\alpha$ doublet, and the broad line. This figure depicts the best fit in the $6.0-6.8~\mathrm{keV}$ energy range. Panel (a) displays the residuals of the best fit. Panel (b) shows the residuals when fitting the line with two Gaussians instead of three. 
{Alt text: Figure of the fitted spectrum, and the ratio of the difference between data and model and the error of each data point. The figure shows the best fit found with three Gaussians, each depicted separately. The continuum below the line is also shown. The residuals of the fit with two Gaussians show significant deviation between data and model, indicating the need for a broad line.} 
 \label{fig:FeKline_3gauss}}
\end{figure}

Fig. \ref{fig:FeKline_3gauss} depicts the best fit to the Fe K$\alpha$ line profile using three Gaussian profiles, with some physical constraints. We set the normalization of the Gaussian representing the Fe K$\alpha_2$ line to be half of that of the Fe K$\alpha_1$ line, to have the same redshift, and the same width. We set the non-redshifted central energies of the Fe K$\alpha_1$ and Fe K$\alpha_2$ peaks to $6.40401~\mathrm{keV}$ and $6.39103~\mathrm{keV}$, respectively, based on the laboratory measurements \citep[Table IV]{1997PhRvA..56.4554H}. Similarly, we used the K$\alpha_1$ to K$\alpha_2$ line ratio to set the non-redshifted peak of the broad line to $6.39968~\mathrm{keV}$\citep[Table V]{1997PhRvA..56.4554H}. The parameters of the Gaussian describing the broad line were allowed to vary freely. 

After subtracting the natural line broadening \citep{1997PhRvA..56.4554H}, the two narrow components were fit with $\sigma=4.3\pm0.2~\mathrm{eV} = (2.0\pm0.1)\times10^2~\mathrm{km}~\mathrm{s}^{-1}$. This corresponds to a $\mathrm{FWHM} =(4.8\pm0.2)\times10^2~\mathrm{km}~\mathrm{s}^{-1}$.

Fig. \ref{fig:FeKline_3gauss} also depicts the residuals when fitting the line with just two Gaussian models, in panel (b). This demonstrates that a broad line is required, and that an underestimated continuum cannot account for it. The broad line was best fit with $\sigma=39\pm3~\mathrm{eV} = (1.8\pm0.1)\times10^3~\mathrm{km}~\mathrm{s}^{-1}$. This corresponds to $\mathrm{FWHM} = (4.3\pm0.3)\times10^3~\mathrm{km}~\mathrm{s}^{-1}$. There are no IR lines with comparably large FWHMs.

The Gaussian describing the narrow Fe K$\alpha_1$ line has a flux of $1.12^{+0.05}_{-0.07}\times10^{-12}~\mathrm{erg}~\mathrm{cm}^{-2}~ \mathrm{s}^{-1}$, an equivalent width of $33\pm2~\mathrm{eV}$, and is redshifted by $(1.99\pm0.03)\times10^{-3}$, which corresponds to $(5.97\pm0.08)\times10^2~\mathrm{km}~\mathrm{s}^{-1}$.  Due to the constraints of the fit, the Fe K$\alpha_2$ line has half of the flux, equivalent width, and the same redshift. The broad line has a flux of $(1.6\pm0.1)\times10^{-12}~\mathrm{erg}~\mathrm{cm}^{-2}~ \mathrm{s}^{-1}$, an equivalent width of $(46\pm3)~\mathrm{eV}$, and is redshifted by more than the narrow peaks, to $(2.4\pm0.3)\times10^{-3}$, or $(7.3\pm0.8)\times10^2~\mathrm{km}~\mathrm{s}^{-1}$. This is likely because the line profile is not symmetric and may possess a Compton shoulder at low energies. A Compton shoulder appears as a broad extension of the low-energy line profile. However, the observed broad line extends towards both higher and lower energies, so it cannot be solely attributed to the Compton shoulder. As the redshift of the broad line modeled with a Gaussian is only slightly higher than that of the narrow core, we find that the Compton shoulder likely only contributes a small fraction to the flux of the broad line. This indicates that the line is produced by a Compton-thin emitter. 

The total flux of the Fe K$\alpha$ line found by this model, is $(3.2\pm0.2)\times10^{-12}~\mathrm{erg}~\mathrm{cm}^{-2}~ \mathrm{s}^{-1}$, which corresponds to an equivalent width of $96\pm4~\mathrm{eV}$. This is large compared to recent measurements of the line flux in other observations \citep{2024ApJ...961..150B, 2024PASJ...76..923I}. This might indicate that previous measurements underestimated the flux of the broad line. Alternatively, this could be partially attributed to a temporarily reduced continuum flux. However, it is not the largest equivalent width that has been measured for the line, as a few previous observations found even larger values \citep{2011ApJ...733...23R, 2016ApJ...821...15F}. 

This best fit of the $6.0-6.8~\mathrm{keV}$ band with this model has a C-statistic of 1695.87, with 11 free parameters. Compared to the BIC of the best fit using model A (see Section \ref{sec:Modelfreeq}), this has a $\Delta \mathrm{BIC} = 23.77$. This model is a reasonable phenomenological description of the Fe K$\alpha$ line profile. However, it underestimates the strength of the Fe K$\alpha_2$ line, due to the limitations of fitting the complex line shape with a set of Gaussians. There are also several discrepancies between the data and the model at the transition between the intervals where the narrow and the broad components dominate, as well as at the outer ends of the line (see Fig. \ref{fig:FeKline_3gauss}).

\citet{2011ApJ...727...19F} calculated that the equivalent width ($\mathrm{EW}$) of the Fe K$\alpha$ line is approximately related to the Hydrogen column density ($N_{\rm H}$), and the covering factor ($C$) via $\mathrm{EW} \approx \frac{300 ~ C ~ N_H}{4\times10^{23}~\mathrm{cm}^{-2}}~\mathrm{eV}$ for solar abundances of  Fe.  So if we assume a typical Hydrogen column density for Cen A, of $N_{\rm H} \approx 1.7\times10^{23}~\mathrm{cm}^{-2}$ \citep{2024ApJ...961..150B}, the measured equivalent width of the line would require a covering factor of $C\approx 75\%$. 

\subsection{Fitting with one MYTorusL component and an unconstrained $q$}\label{sec:Modelfreeq}

We also considered more physically motivated models. In the following, we fit the Fe K$\alpha$ line using a H{\"o}lzer profile, an analytic approximation to laboratory measurements of the  Fe K$\alpha$ line complex \citep{1997PhRvA..56.4554H}, as implemented in the MYTORUS model \citep{2009MNRAS.397.1549M, 2024MNRAS.527.1093Y}, which also includes the Compton shoulder. The combined spectral model of \texttt{rdblur * atable\{mytl\_V000HLZnEp000\_v01.fits\}} is a full physical model of disk reflection, using the most up-to-date measurements of the FWHM of the Fe K$\alpha_1$, Fe K$\alpha_2$, and Fe K$\beta$ lines, as well as their lab-based intrinsic line profiles. It also uses a base resolution of $2~\mathrm{eV}$, sufficient for an analysis of XRISM/Resolve data. For simplicity, we will refer to the table model \texttt{atable\{mytl\_V000HLZnEp000\_v01.fits\}} as \texttt{MYTorusL}. The line shape produced by \texttt{MYTorusL} does not include any kinematics. The \texttt{rdblur} model blurs the \texttt{MYTorusL} line, corresponding to an emission region between radii of $R_{\rm in}$ and $R_{\rm out}$, for a particular inclination ($i$), and emissivity index ($q$). 

The \texttt{rdblur*MYTorusL} model is the most physically-accurate currently available description of the Fe K$\alpha$ line emitted by an AGN at a sufficient energy resolution. Nevertheless, there are drawbacks of this combination of models, and we use it for lack of a more realistic model. The assumed toroidal structure of the MYTORUS models may bias some of the fitted spectral parameters. \citet{2024MNRAS.527.1093Y} notes that the \texttt{MYTorusL} model can be crudely used to represent other geometries as well, if utilized in a decoupled mode \citep{2012MNRAS.423.3360Y}, as we do. However, the \texttt{rdblur} model assumes that the pre-convolved line shape is emitted at a single radius. The information about the location from which the Fe K$\alpha$ photons were emitted in the \texttt{MYTorusL} model is lost. These inconsistencies may bias some results and offset some of the best-fit spectral parameters from their true values. Therefore, we caution from a too strict interpretation of the results.


We initially fitted the line with a single \texttt{MYTorusL} component, blurred by \texttt{rdblur}. We allowed the inner radius, $R_{\mathrm{in}}$, outer radius, $R_{\mathrm{out}}$, inclination, $i$, and emissivity index, $q$, to vary freely. The full XSPEC model we used for this was: \texttt{powerlaw + ztbabs * (zpowerlw + gauss + gauss + constant * rdblur * MYTorusL)}. We will refer to this as model A. The \texttt{powerlaw} component describes the contaminating emission from other sources. Its parameters were set equal to the best-fit values found in the Chandra spectrum. The \texttt{zpowerlw + gauss + gauss} components describe the continuum around the line. Their best-fit parameters were constrained within the $1\sigma$ limits of the fit to the $5.8-7.6~\mathrm{keV}$ band, except for the normalizations, which were allowed to vary freely. All redshifts were tied together, and this single value was allowed to vary freely. All parameters of the \texttt{MYTorusL} line were tied to those of the \texttt{zpowerlaw} component, and the equatorial $N_{\rm H}$ was allowed to vary freely. The \texttt{constant} parameter, $k$ in Table \ref{tab:bestfitpropsFeline},  describes the relative normalization, which accounts for a variety of model dependencies such as the delayed response between the continuum and the line, possible non-solar abundances or geometries not consistent with the assumptions of the \texttt{MYTorusL} and \texttt{rdblur} models, such as the covering factor or radial dependency of the density.  

The MYTORUS scattered continuum component was not included in this model, as it only has a minimal impact on the spectrum shape in the energy band of $6.0 - 6.8~\mathrm{keV}$. Within this energy range, and for the measured equatorial $N_{\rm H}$ values, including it only has the effect of requiring a slightly smaller \texttt{zpowerlw} normalization to fit the data. We used \texttt{zpowerlw} to describe both the zeroth order and scattered continuum. We note that $q$ is described by the parameter \texttt{Betor10} within the \texttt{rdblur} model, which is equal to $-q$. 

\begin{figure}[t]
\begin{center}
\resizebox{\hsize}{!}{\includegraphics{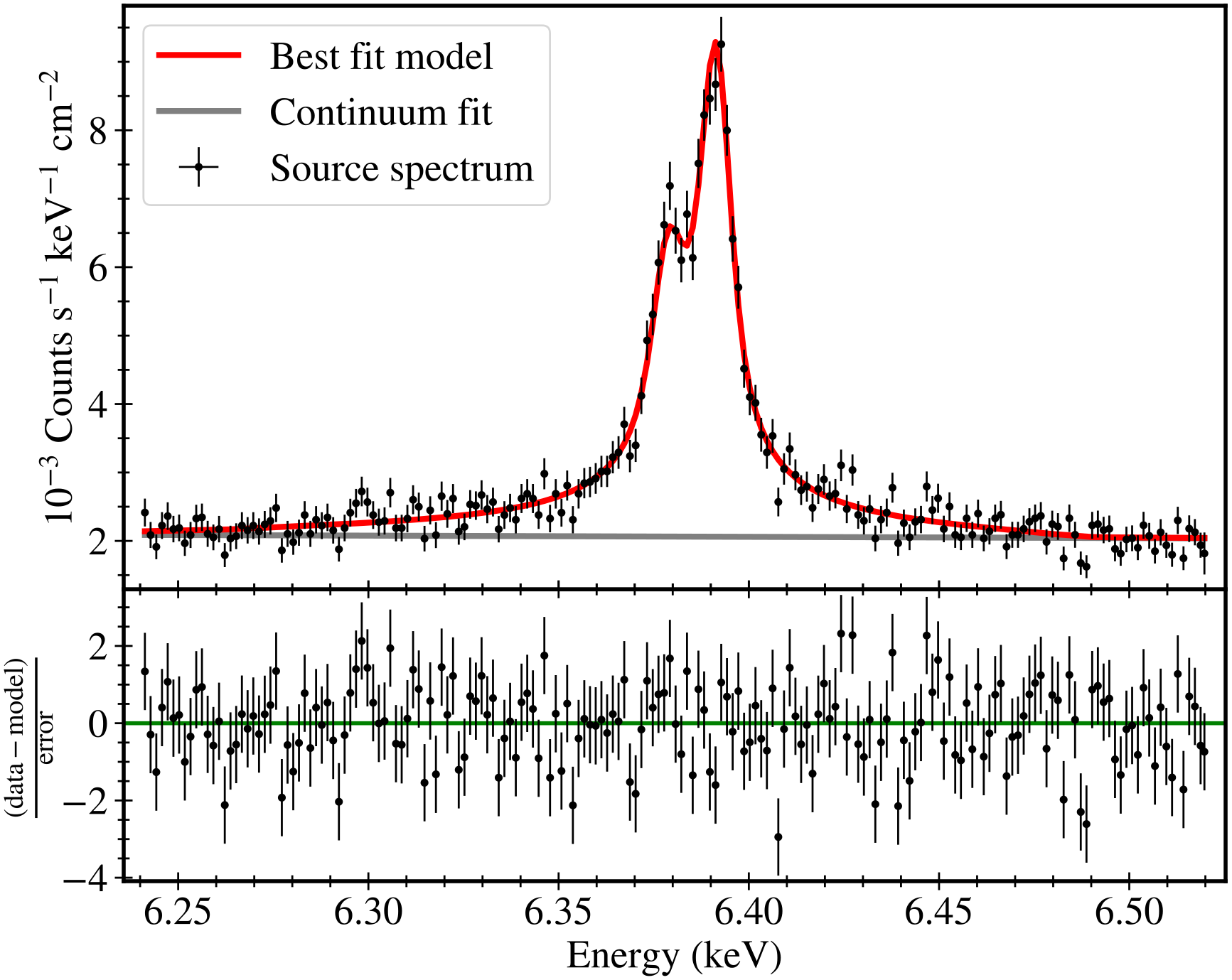}}
\end{center}
\caption{XRISM/Resolve spectrum of the Fe K$\alpha$ line, fitted with model A, featuring one \texttt{MYTorusL} line model, an emissivity index that is free to vary, and a constant normalization scaling factor. This figure depicts the best fit in the $6.24-6.52~\mathrm{keV}$ energy range. The spectrum was fitted over a wider energy range of $6.0-6.8~\mathrm{keV}$. This and the following figures show a spectrum that has been rebinned for visual clarity only. 
{Alt text: Figure of the fitted spectrum, and the ratio of the difference between data and model and the error of each data point. The figure shows the best fit found with model A. The continuum below the line is also shown.} 
 \label{fig:FeKline_1MYTfreeq}}
\end{figure}

The best fit of the spectrum using model A is shown in Fig. \ref{fig:FeKline_1MYTfreeq}, and its parameters are listed in Table \ref{tab:bestfitpropsFeline}. It provides a very good description of the line profile ($\mathrm{C}=1664.72$) with minimal free parameters ($\mathrm{DOF}=1587$). It has the lowest BIC of any model we investigated ($1753.25$). 

However, this fit required a relative normalization parameter of $k = 4.1^{+6.0}_{-1.3}$. The cause of the large uncertainty in this factor is the degeneracy between it and the equatorial $N_{\rm H}$ used by \texttt{MYTorusL}. However, the line profile cannot be adequately reproduced with a multiplicative constant of $k=1$, as a larger equatorial $N_{\rm H}$ cannot fully account for both the shape and the strength of the line. 

\begin{figure}[t]
\begin{center}
\resizebox{\hsize}{!}{\includegraphics{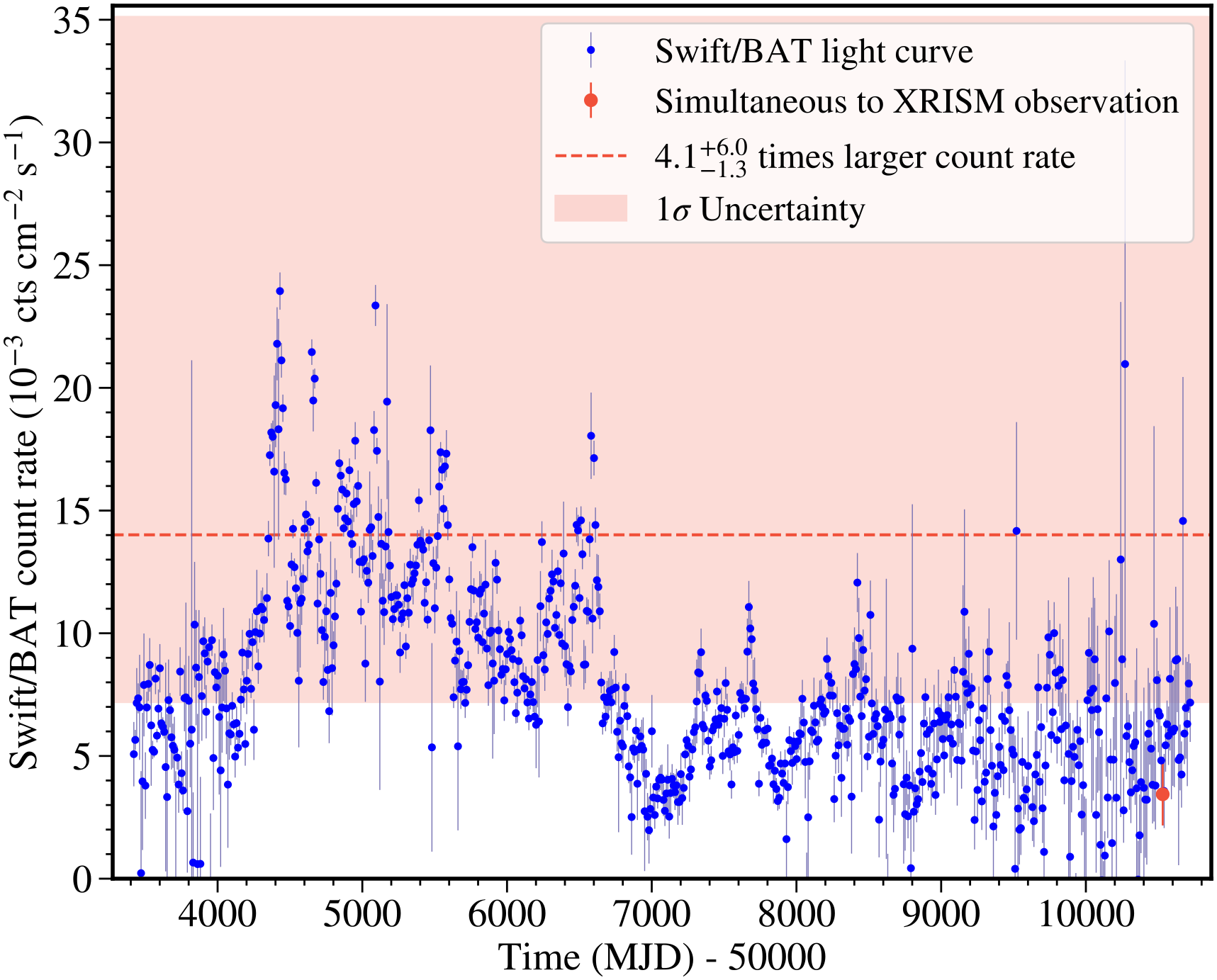}}
\end{center}
\caption{Swift/BAT light curve of Cen A. Each data point represents the average count rate over a 10-day interval. The count rate measured simultaneously to the XRISM observation is indicated in red. The $4.1^{+6.0}_{-1.3}$ multiple of that is depicted as a dashed line, and the pink region indicates the $1\sigma$ error. 
{Alt text: Figure showing the light curve of Cen A as observed by Swift/BAT. There is a lot of variation, but Cen A initially became brighter, before gradually becoming fainter, and subsequently varying around a constant low count rate level. A large pink rectangle encompasses the $1\sigma$ error of the $4.1^{+6.0}_{-1.3}$ multiple of the measured count rate simultaneous to the XRISM observation. It includes most measurements, but not as many more recent results. The box extends far above the highest count rate measurement.} 
 \label{fig:SwiftBATlc}}
\end{figure}

We investigated to what extent this factor could be accounted for by the variability of the X-ray emission of Cen A, if we assume that it is entirely due to lags between the continuum and the line emission. Fig. \ref{fig:SwiftBATlc} depicts the long-term hard X-ray light curve of Cen A observed by Swift/BAT. The continuum was indeed somewhat fainter during the XRISM observation than in previous weeks. The ratio of the average Swift/BAT count rate measured prior to the XRISM observation, to the Swift/BAT count rate measured in a 10-day bin simultaneous to it, is $2.3\pm0.8$. The large $1\sigma$ uncertainties of both this ratio and the best-fit relative normalization overlap, but the best-fit value of $k$ is larger. 

There are other factors that could help to explain the large best-fit value of $k$. These include non-solar abundances, variable solid angles, or reduced absorption of the Fe K$\alpha$ line photons compared with the continuum. A geometry that differs to that assumed by the \texttt{MYTorusL} and \texttt{rdblur} models could also contribute. We used \texttt{zpowerlw} to describe both the zeroth order and scattered continuum. The scattered continuum only contributes a small fraction of the power-law flux. When distinguishing the two components, the value of $k$ is minutely increased, as it is measured relative to the zeroth-order continuum. 

The best-fit inclination of this model is $24^{+13}_{-7}~\degree$. This inclination is mainly constrained by the observed shape of the broad tails of the line profile. Low inclinations are disfavored. For example, a $10\degree$ inclination is excluded at a $7.2 \sigma$ level.  In contrast, very high inclinations only result in slightly worse fits. Although the $1\sigma$ upper bound on the inclination is at $37\degree$, an inclination of $90\degree$ is only disfavored at a $1.6\sigma$ level. 

There is a strong degeneracy between the best-fit values of the inner and outer radii and the inclination. Nevertheless, the fit indicates that the Fe K$\alpha$ line is emitted from a wide range of different radii from the black hole. The best-fit values of $R_{\rm in} = 5.4^{+7.8}_{-4.0}\times10^2~r_{\rm g} = 1.4^{+2.1}_{-1.1}\times10^{-3}~\mathrm{pc}$, and $R_{\rm out} = 6.6^{+13.8}_{-3.2}\times10^6~r_{\rm g} = 17^{+36}_{-9} ~\mathrm{pc}$ indicate that the line could be produced in an optically invisible broad line region (BLR), the torus, and potentially also the molecular disk. For converting between gravitational radii and parsecs, we used the measured mass of the supermassive black hole in Cen A, of $5.5\pm3.0\times10^7~\mathrm{M}_{\odot}$  \citep{2009MNRAS.394..660C, 2022ApJS..261....2K}. That uncertainty describes the $3\sigma$ error. As we are using $1\sigma$ errors throughout this paper, we assume the $1\sigma$ errors are $\pm1.0\times10^7~\mathrm{M}_{\odot}$.

The best-fitting emissivity index was found to be $q=1.99\pm0.03$. This is tightly constrained and inconsistent with a value of $q=3$. That instance will be explored in more detail in Section \ref{sec:FeKq3}. The line is redshifted by $(1.89\pm0.03)\times10^{-3}$. The total flux of the line is $3.5\pm0.2\times10^{-12}~\mathrm{erg}~\mathrm{cm}^{-2}~ \mathrm{s}^{-1}$. 

\subsection{Fitting with two \texttt{MYTorusL} components with an unconstrained $q$}\label{sec:Model2comp}

We investigated whether additional complexity in the spectral model could improve the ability to fit the line profile. We also sought to determine whether the large relative normalization factor required by model A could instead be accounted for by a more complex geometry. 

\begin{figure}[t]
\begin{center}
\resizebox{\hsize}{!}{\includegraphics{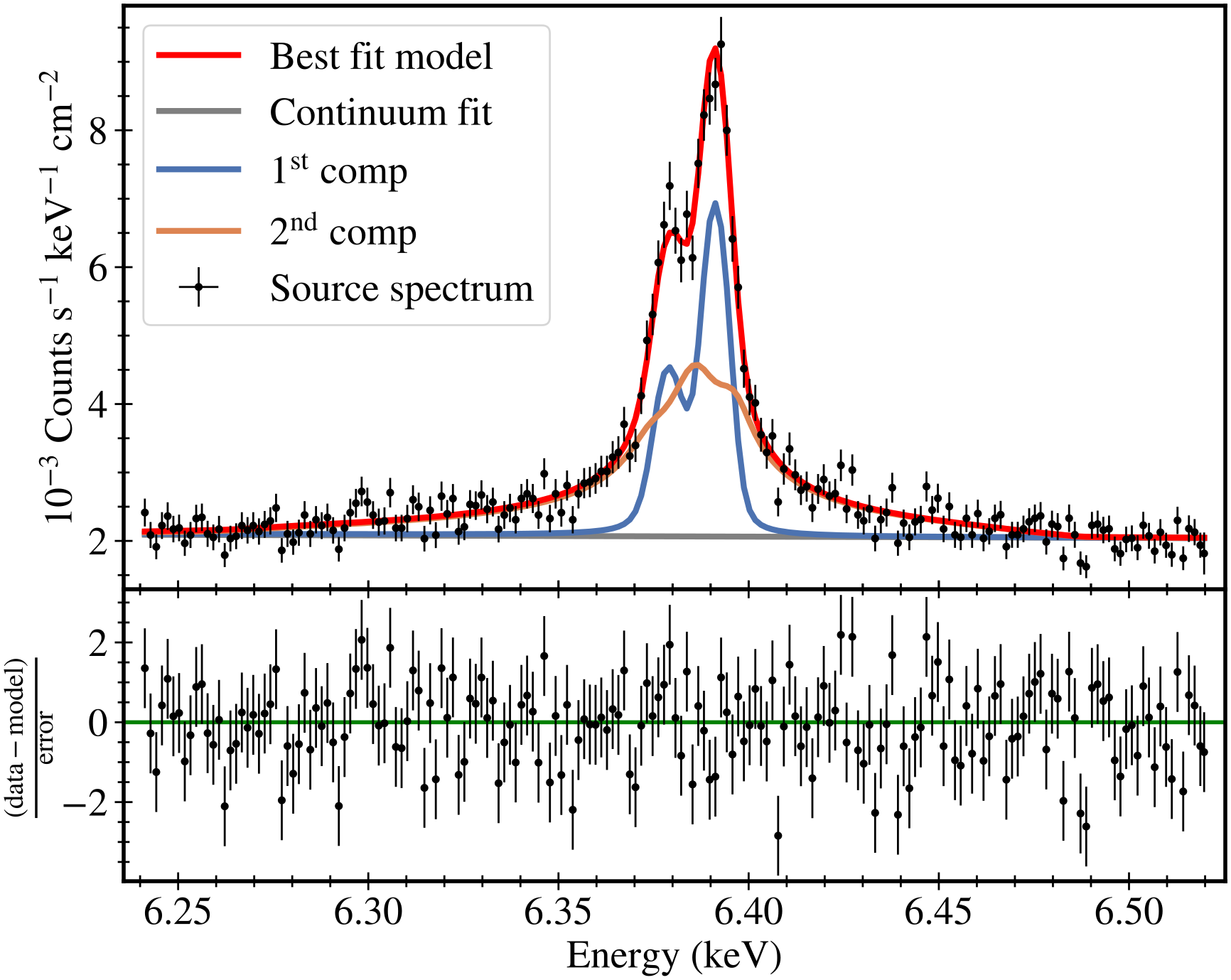}}
\end{center}
\caption{XRISM/Resolve spectrum of the Fe K$\alpha$ line, fitted with model B;  two \texttt{MYTorusL} line models and an emissivity index that is free to vary. 
{Alt text: Figure of the fitted spectrum, and the ratio of the difference between data and model, and the error of each data point. The figure shows the best-fit found with model B, and also depicts the two blurred line components that contribute to it. The continuum below the line is also shown.} 
 \label{fig:FeKline_2MYT}}
\end{figure}

Therefore, we also fitted the Fe K$\alpha$ line with two blurred \texttt{MYTorusL} components. There is a lot of freedom to fit the line profile with two \texttt{MYTorusL} models, and several different solutions are possible. One way of fitting the data is to let one component fit the narrow double-peaks, and the second component fit the broad line. We will explore another approach to fitting the line profile with two \texttt{MYTorusL} models in Appendix \ref{sec:4MYT}.

When fitting this model, we found that the outer radius of the second line component, which produces the broad line, was consistent with the inner radius of the first component. Similarly, both components had consistent best-fit inclinations and emissivity indices. Therefore, we tied these parameters together, as there was insufficient evidence that they are independent. 

We will refer to this as model B, and it is defined, in XSPEC notation, as \texttt{powerlaw + ztbabs * (zpowerlw + gauss + gauss + constant * rdblur * MYTorusL + constant * rdblur * MYTorusL)}. The difference between model B and model A, is that this one can account for different equatorial Hydrogen column densities at different radii. It also allows us to determine whether the best-fit emissivity index found for model A is a consequence of the assumptions of that model. Additionally, the Compton shoulder produced by two weaker line components may differ from the Compton shoulder of a single stronger line.    

The best fit to the line profile using model B is indicated in Fig. \ref{fig:FeKline_2MYT}, and the fitting parameters are listed in Table \ref{tab:bestfitpropsFeline}. The best-fit parameters with model B are consistent with those found for model A. The emissivity index has a best-fit value of $2.08^{+0.06}_{-0.07}$, and the inclination is $23^{+18}_{-4}~\degree$. The extra freedom to fit the data with model B enables the C-statistic to be reduced by $2.93$ compared with the best fit with model A. However, it requires 3 additional free parameters, so the best-fit with model B has a BIC that is larger by $19.22$. 

The degeneracy between the two relative normalizations and the two Hydrogen column densities of model B means that none of these four parameters is well-constrained. The line component flux is defined by the relative normalization and the Hydrogen column density. The emissivity index blurs each line component individually, and does not reduce the line flux of the outer component accordingly. This means that the reduction in emissivity with radius between the first and second component is instead accounted for by a reduced $N_{\rm H}$ or relative normalization factor for the outer component. This means that best-fit values of $N_{\rm H}$ and $k$ of the outer line component should not be regarded as accurate representations of their physical values. To extract accurate physical values for these parameters would require an investigation of the exact interaction of the $N_{\rm H}$, $k$, $R_{\rm in}$, $R_{\rm out}$, and $q$ parameters in shaping the amplitude of the outer line component. The large uncertainties on many of these parameters prevented us from determining well-constrained, physically accurate values for $N_{\rm H}$ and $k$ of the outer component.   

The radius at which the inner component ends, and the lower-density outer component starts, is $3.5^{+4.7}_{-2.7}\times10^5~r_{\rm g}$, which corresponds to $0.91^{+1.24}_{-0.72}~\mathrm{pc}$. The two relative normalization parameters have large uncertainties, but best-fit values of $2.5$ and $2.6$, which are easier to justify based on the known variability of the continuum. The range of radii that contribute to the Fe K$\alpha$ line again range from $5.4^{+10.4}_{-2.2}\times10^2~r_{\rm g} = 1.4^{+2.8}_{-0.6}\times10^{-3}~\mathrm{pc}$ to $1.7^{+9.5}_{-0.8}\times10^6~r_{\rm g} = 4.5^{+25.0}_{-2.4}~\mathrm{pc}$. This outermost radius is resolvable with ALMA or JWST data. 

\begin{table*}
\centering
  \tbl{Best fit properties of the Fe K$\alpha$ line models}{
  \addtolength{\tabcolsep}{-0.2em}
  \renewcommand{\arraystretch}{1.3}
  \begin{tabular}{ccc|cccc}
      \hline
      Component & & Units & Model A & Model B & Model C & Model D  \\ \hline
      \texttt{constant} & $k$ & & $4.1^{+6.0}_{-1.3}$ & $2.6^{+13.0}_{-2.0}$ & $1.0$ \footnotemark[$\S$] & $1.0$ \footnotemark[$\S$] \\ \hline
      \texttt{rdblur} \footnotemark[$*$] & $q$ & & $1.99\pm0.03$ & $2.08^{+0.06}_{-0.07}$ & $2.2\pm0.2$ & $3$ \footnotemark[$\S$]  \\
       & $R_{\rm in}$ & $r_{\rm g}$ & $5.4^{+7.8}_{-4.0}\times10^2$ & $3.5^{+4.7}_{-2.7}\times10^5$ & $7.2^{+5.6}_{-0.9}\times10^5$  & $2.4^{+1.2}_{-0.7}\times10^5$ \\
       & $R_{\rm out}$ & $r_{\rm g}$ & $6.6^{+13.8}_{-3.2}\times10^6$ & $1.7^{+9.5}_{-0.8}\times10^6$ & $7.9\times10^5$ \footnotemark[$\|$] & $8.3\times10^8$ \footnotemark[$\ddag$] \\
       & $i$ & $\deg$ & $24^{+13}_{-7}$ & $23^{+18}_{-4}$ & $30^{+12}_{-3}$ & $23\pm4$ \\ \hline
      \texttt{MYTorusL} & $N_{\rm H}$ & $10^{24}~\mathrm{cm}^{-2}$ & $0.10^{+0.06}_{-0.10}$ & $5.8^{+34.7}_{-5.8}\times10^{-2}$ & $0.25^{+0.03}_{-0.02}$ & $0.24^{+0.03}_{-0.05}$ \\ 
       & $z$ & $10^{-3}$ & $1.89\pm0.03$ & $1.89\pm0.03$ & $1.89\pm0.03$ & $1.88\pm0.03$ \\ \hline
      \texttt{constant} & $k$ & & & $2.5^{+19.6}_{-1.4}$ & $1.0$ \footnotemark[$\S$] & $1.0$ \footnotemark[$\S$] \\ \hline
      \texttt{rdblur} & $R_{\rm in}$ & $r_{\rm g}$ & & $5.4^{+10.4}_{-2.2}\times10^2$ & $5.7^{+4.5}\times10^4$ & $1.2^{+0.7}_{-0.4}\times10^4$ \\ 
      & $R_{\rm out}$ & $r_{\rm g}$ & & $3.5\times10^5$ \footnotemark[$\dag$] &  $6.3\times10^4$ \footnotemark[$\|$] & $2.4\times10^5$ \footnotemark[$\dag$]  \\ \hline
      \texttt{MYTorusL} & $N_{\rm H}$ & $10^{24}~\mathrm{cm}^{-2}$ & & $0.11^{+0.24}_{-0.11}$ & $6.9^{+1.6}_{-3.2}\times10^{-2}$ & $0.12^{+0.03}_{-0.01}$ \\ \hline
      \texttt{rdblur} & $R_{\rm in}$ & $r_{\rm g}$ & & & $1.0^{+0.8}_{-0.2}\times10^{3}$ & $8.3^{+15.4}_{-2.1}\times10^2$ \\ 
       & $R_{\rm out}$ & $r_{\rm g}$ & & & $3.3_{-1.0}\times10^4$ & $1.2\times10^4$ \footnotemark[$\dag$] \\ \hline
      \texttt{MYTorusL} & $N_{\rm H}$ & $10^{24}~\mathrm{cm}^{-2}$ & & & $0.17^{+0.12}_{-0.02}$ & $0.11^{+0.02}_{-0.01}$ \\ \hline
       & C & & $1664.72$ & $1661.81$ & $1661.03$ & $1668.12$ \\
       & DOF & & $1587$ & $1584$ & $1584$ & $1585$ \\
       & $\Delta \mathrm{BIC}$ & & $0$ & $19.22$ & $18.44$ & $18.15$ \\ \hline
    \end{tabular}}\label{tab:bestfitpropsFeline}
\begin{tabnote}
\footnotemark[] All models fit the energy range $6.0-6.8~\mathrm{keV}$, containing 1599 bins of the XRISM/Resolve spectrum. Model A fits the Fe K$\alpha$ line with one \texttt{MYTorusL} component, a multiplicative factor, and an unconstrained $q$. Model B extends model A to two \texttt{MYTorusL} components. Model C consists of three \texttt{MYTorusL} lines, uses a free $q$, and sets the outer radius of the first and second components equal to $1.1~R_{\rm in}$. The inner radii of the outer two components were bounded by the radii found from reverberation mapping, within $1\sigma$ errors from fitting the lags and the measured black hole mass. Model D consists of three \texttt{MYTorusL} lines, each of which has $q=3$, the same inclination, redshift, and corresponding inner and outer radii. If the number listed only includes an upper error, then the lower bound is not constrained. The same applies when the upper error is missing. The $\Delta \mathrm{BIC}$ is stated relative to the best-fit with model A. \\
\footnotemark[*] The emissivity and inclination of all additional \texttt{rdblur} components is set equal to the values of this one.  \\
\footnotemark[\dag] The outer radius of these components is set equal to the inner radius of the previous component, in models A, C, and D. \\
\footnotemark[\S] This is a fixed quantity.  \\
\footnotemark[\ddag] Neither the upper nor the lower bound of this parameter could be determined. \\
\footnotemark[$\|$] This is set to be $1.1\times R_{\rm in}$.
\end{tabnote}
\end{table*}

\subsection{Radii defined by reverberation mapping}\label{sec:revmapspec}

\citet{2024PASJ...76..923I} carried out reverberation mapping studies of the delay between variations of the power law emission and the Fe K$\alpha$ line in Cen A. They found that the lag between these two spectral features could be explained by the existence of two Fe K$\alpha$ line emitting regions; one at a radius of $0.19^{+0.10}_{-0.02}~\mathrm{pc}$, corresponding to $230^{+120}_{-20}$ light days and one at a radius of $>1.7~\mathrm{pc}$ or $>5.5$ light years. These radii correspond to $7.2^{+4.0}_{-1.5}\times10^4~r_{\rm g}$, and $>(6.5\pm1.2)\times10^5~r_{\rm g}$. 

We investigated to what extent this physical description found from reverberation mapping could be applied to fit the profile of the Fe K$\alpha$ line. To do so, we defined a spectral model similar to that of model B, but with the inner radius of the inner component constrained to within $5.7~{\rm and}~ 11.2\times10^4~r_{\rm g}$, and the inner radius of the outer component constrained to $>5.3\times10^5~r_{\rm g}$. Furthermore, we specified that the outer radii of the two components could only be $10\%$ larger than the inner radii, as significantly extended emission regions would not result in two specific radii found from lag measurements in reverberation mapping. We also allowed the emissivity index to vary freely. When fitting the data with this model, we found that the central peaks of the line are well described, but the broad tails are significantly underestimated. This remained true even for fits with high inclinations. 

Therefore, we explored whether the addition of a third \texttt{MYTorusL} component to the model, with an inner radius of $<5.7\times10^4~r_{\rm g}$, could allow us to fit the line profile. We found that a model consisting of three \texttt{MYTorusL} lines with identical emissivity indices and inclinations provided a very good fit to the data. As there are three \texttt{MYTorusL} components, there is a huge degeneracy between the six parameters of the relative normalizations and the $N_{\rm H}$ of each. None of these parameters can be adequately constrained. Instead, we set all relative normalization parameters to 1. This had no impact on the best fit to the data, except that the individual best-fit $N_{\rm H}$ values are possibly slightly too large, and have much smaller uncertainties than they should have. Furthermore, as discussed in Section \ref{sec:Model2comp}, the $N_{\rm H}$ values of the outer two components are not directly related to the physical values of these parameters in multi-component models. We refer to this as model C. In XSPEC notation, it is described as: \texttt{powerlaw + ztbabs * (zpowerlw + gauss + gauss + rdblur * MYTorusL + rdblur * MYTorusL + rdblur * MYTorusL)}.

\begin{figure}[t]
\begin{center}
\resizebox{\hsize}{!}{\includegraphics{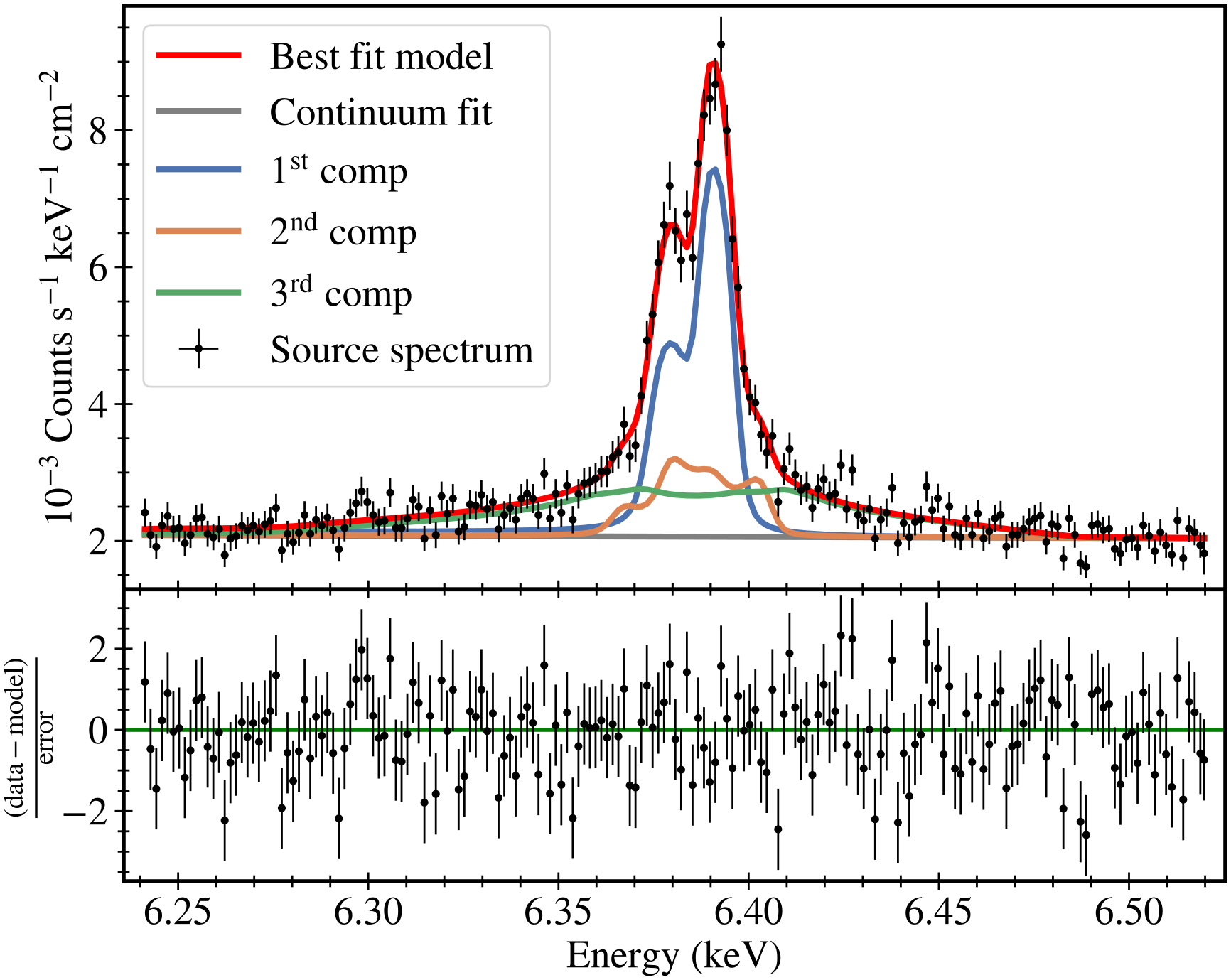}}
\end{center}
\caption{XRISM/Resolve spectrum of the Fe K$\alpha$ line, fitted with model C; three \texttt{MYTorusL} components and an emissivity index that is free to vary, but the radii set equal to those found from reverberation mapping by \citet{2024PASJ...76..923I}. 
{Alt text: Figure of the fitted spectrum, and the ratio of the difference between data and model and the error of each data point. The figure shows the best-fit found with model C, and also depicts the three blurred line components that contribute to it. The continuum below the line is also shown.}  
 \label{fig:FeKline_Iwata}}
\end{figure}

The best fit with model C is comparable to that with model B. It is depicted in Fig. \ref{fig:FeKline_Iwata}, and the best-fitting parameters are listed in Table \ref{tab:bestfitpropsFeline}. The C-statistic is reduced by $0.78$ compared to model B, for the same number of degrees of freedom. The third \texttt{MYTorusL} component is fitted to have inner and outer radii of $1.0^{+0.8}_{-0.2}\times10^{3} ~ r_{\rm g} = 2.6^{+2.2}_{-0.7} \times10^{-3}~\mathrm{pc}$, and $3.3_{-1.0}\times10^4 r_{\rm g} = 8.7_{-3.1}\times10^{-2} ~\mathrm{pc}$, respectively. The $1\sigma$ upper limit of the outer radius could not be determined. It is challenging to identify this component through reverberation mapping, since it would require monitoring on timescales of weeks, and would produce a range of lags, rather than a single value. The reverberation mapping analysis by \citet{2024PASJ...76..923I} is unlikely to be sensitive to this component, as they rebinned the Swift/BAT light curve into 20-day intervals and the Fe K$\alpha$ line flux data have large time gaps, ranging from around a hundred to a thousand days. 

We conclude that the spectral shape of the Fe K$\alpha$ line is consistent with the radii measured by reverberation mapping, if a third emission region closer to the black hole is included in the model. However, the relative fluxes of these three line components found from spectral modelling differ from those derived from reverberation mapping. \citet{2024PASJ...76..923I} found that the intermediate component, at a radius of $0.19^{+0.10}_{-0.02}~\mathrm{pc}$ contributed $56-86\%$ of the total line flux. In contrast, we found that it contributes $16^{+3}_{-12}\%$, whereas the outer component contributes $49\pm4\%$, and the inner extended component contributes $37^{+37}_{-4}\%$. The different relative strengths could, however, partially be due to a decreasing continuum flux in the years preceding the observation. 

The constraints we placed on the radii of the outer two components result in a best-fit inclination of $30^{+12}_{-3}~\degree$. Unlike in models A, and B, high inclinations are strongly ruled out in model C, due to the constraints on the radii. An inclination of $90\degree$ is disfavoured at a $6.7\sigma$ level. The best-fit emissivity index of model C is $q=2.2\pm0.2$, consistent with the results of models A and B. 

\subsection{Fitting with $q=3$}\label{sec:FeKq3}

An emissivity index of $q=3$ is often assumed in spectral analysis. It corresponds to the emissivity profile expected for a lamppost corona at radii of $\gtrsim 30~r_{\rm g}$ illuminating a disk. In this subsection, we will also investigate how to fit the Fe K$\alpha$ line with this assumption. 

A H{\"o}lzer line profile \citep{1997PhRvA..56.4554H}, blurred by the \texttt{rdblur} model with an emissivity power-law index of $q=3$ has less bright narrow lines for the same broad-line flux, when compared with $q=2$. The observed ratio of broad to narrow line strengths in the XRISM/Resolve Cen A spectrum more closely matches the expectation from $q=2$, which is why the best-fit emissivity index in models A, B, and C was $q\approx2$.

A single \texttt{MYTorusL} component with $q=3$ does not fit the observed Fe K$\alpha$ line profile well. The inclusion of a second \texttt{MYTorusL} line model significantly improves the quality of the fit, via a reduction of the BIC by $197$. This second component is closer to the black hole and describes the broad line. However, even this model fails to adequately describe the full Fe K$\alpha$ line profile, which is why $q=3$ is strongly ruled out in the best-fit with model B. A two-component model with $q=3$ still underestimates the outer edges of the line profile, while overestimating the central parts. We found that three \texttt{MYTorusL} components are required to describe the Fe K$\alpha$ line profile with an accuracy comparable to that achieved by the best fits of models A, B, and C, while also requiring $q=3$. This third component would be the closest to the black hole, and would describe the outer edges of the Fe K$\alpha$ line profile. The inclusion of this third component further reduces the BIC of the best fit by $56$. 

In XSPEC notation, the full model with three \texttt{MYTorusL} lines, which we will refer to as model D, is written identically to model C. However, the inner radii of all three components are free to vary in model D, whereas the emissivity indices are constrained to $q=3$. We again set the relative normalization constants equal to 1 when fitting the spectrum with this model. As we had done for model B, we set the inclinations and redshifts of all components equal to one another, after finding that they were consistent within errors. We also set the outer radii of the first and second line components equal to the inner radii of the second and third components, respectively. 

\begin{figure}[t]
\begin{center}
\resizebox{\hsize}{!}{\includegraphics{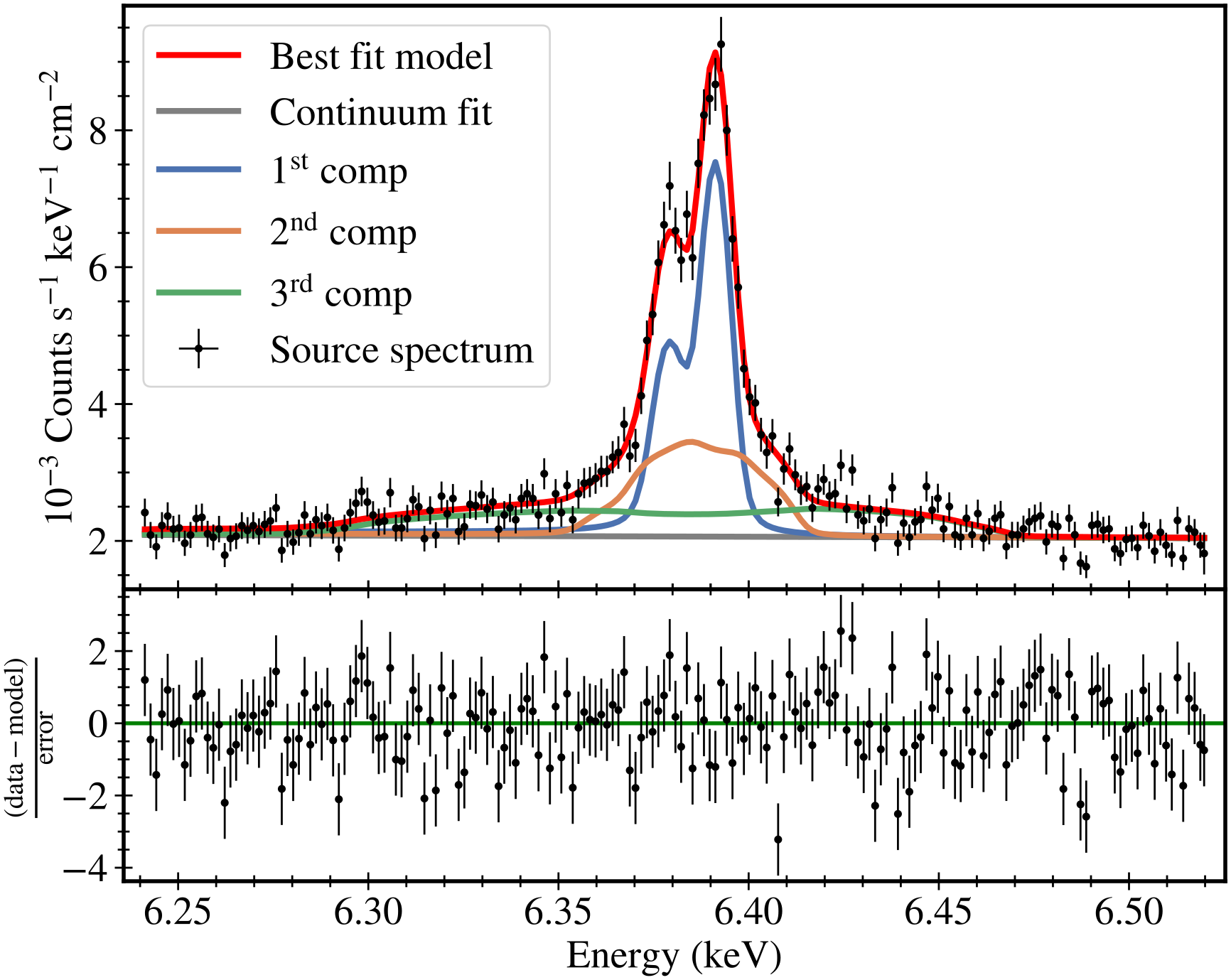}}
\end{center}
\caption{XRISM/Resolve spectrum of the Fe K$\alpha$ line, fitted using model D, which features three blurred \texttt{MYTorusL} components and an emissivity index that is set to $q=3$.
{Alt text: Figure of the fitted spectrum, and the ratio of the difference between data and model and the error of each data point. The figure shows the best-fit found with model D, and also depicts the three blurred line components that contribute to it. The continuum below the line is also shown.}  
 \label{fig:FeKline_3MYT}}
\end{figure}

The best fit with model D is depicted in Fig. \ref{fig:FeKline_3MYT}, and the best-fit parameters are listed in Table \ref{tab:bestfitpropsFeline}. The best fit with model D has a larger C-statistic than the ones with models A, B, or C, but a slightly lower BIC than for models B and C. To compensate for the faster drop in emissivity with radius, this model assigns higher Hydrogen densities to outer regions of the accretion flow, compared with model C. Simplistically, a density that increases with radius can produce a similar line shape with $q=3$, as a uniform density structure with $q=2$. The best-fit inclination of all three line components was found to be $23\pm4\degree$, consistent with the best-fits found by the other models. 

We also investigated whether the addition of a fourth \texttt{MYTorusL} component would be statistically significant compared to model D. A fourth component, however, always resulted in a larger BIC, even when reducing the C-statistic. The component that produced the greatest reduction in C-statistic compared to this model is discussed in Appendix \ref{sec:4MYT}, albeit in addition to model A.

\section{Caveats and Open Questions}\label{sec:caveats}

The \texttt{MYTorusL} and \texttt{rdblur} models are based on several specific assumptions. The results presented here could be biased by them, and these assumptions may or may not be correct. 

The \texttt{MYTorusL} model assumes a torus of line-emitting material, with an opening angle of $60\degree$. However, based on the equivalent width of the line, we estimate a covering factor of $\approx 75\%$. Similarly, \citet{2025arXiv250205002M} detected a giant torus in the system with a large opening angle. Differences between the reprocessor geometry assumed by \texttt{MYTorusL} and its actual shape may affect some parameters, especially the relative normalizations.

Models A to D consistently fit a low inclination. This preference is predominantly due to the outer edge of the line profile. The central parts of the line shape can be reproduced with almost any inclination. We assumed that the narrow and broad line components had the same inclination, but this is not necessarily true. Appendix \ref{sec:4MYT} discusses another solution, which fits the outer wings of the line independently, and thereby allows the central profile to be fitted with a model that cannot distinguish between high or low inclinations. 

The \texttt{rdblur} model assumes a sharp boundary at the inner and outer edges of a cylindrically symmetric disk, as well as a consistent emissivity index. This assumption also drives some of the derived parameters. Absorption in the torus reduces the incident X-ray flux. This was not accounted for in our models, which might mean that the best-fit emissivity indices were slightly overestimated. However, as the equatorial $N_{\rm H}$ was found to be Compton-thin, the likely impact on the measured emissivity is minimal.

The \texttt{rdblur} model assumes that the non-broadened line profile produced by \texttt{MYTorusL} originates from a single radius. However, the line profile of \texttt{MYTorusL} is generated for a non-rotating extended toroidal structure. The information about the location of the emitted Fe K$\alpha$ line photons is lost. An ideal spectral model would immediately apply the kinematic information in situ while calculating a \texttt{MYTorusL}-like line profile. However, such a model is not currently available. The different geometries assumed by the two models, as well as the lost information of the locations at which the Fe K$\alpha$ photons were emitted, may have an impact on the measured physical parameters. However, this impact may be reduced in the Compton-thin regime.

We restricted these spectral fits to a narrow energy range. Constraints provided by future modeling of the shape of the Fe K$\beta$ line and the broad-band continuum may have an impact on these results.

\section{Discussion}\label{sec:disc}

The Fe K$\alpha$ line in Cen A has a complex shape, which can only be explained by contribution from emission regions distributed over a wide range of radii, from $\sim10^{2-3}~r_{\rm g}\sim10^{-3}~\mathrm{pc}$ to $\gtrsim10^{6}~r_{\rm g}$, which corresponds to $\gtrsim10~\mathrm{pc}$. The main challenge of fitting the line is accounting for both the narrow, well-resolved peaks of the Fe K$\alpha_1$ and Fe K$\alpha_2$ lines, as well as the broad line. 

The best fit with the fewest fitting parameters is found for model A. It successfully fits the line profile with a single uniform emission region extending between radii of $5.4^{+7.8}_{-4.0}\times10^2~r_{\rm g} = 1.4^{+2.1}_{-1.1}\times10^{-3}~\mathrm{pc}$ and $6.6^{+13.8}_{-3.2}\times10^6~r_{\rm g} = 17^{+36}_{-9}~\mathrm{pc}$ at an inclination of $24^{+13}_{-7} ~\degree$, and an emissivity index of $q=1.99\pm0.03$. It requires a large relative normalization parameter, which might be justified via a combination of past variability, the low continuum flux at the time of observation, possibly different absorption, as well as non-solar abundances, a differing opening angle, or other geometric differences not accounted for by the model. 

The Fe K$\alpha_1$ and Fe K$\alpha_2$ lines have a $\mathrm{FWHM} =(4.8\pm0.2)\times10^2~\mathrm{km}~\mathrm{s}^{-1}$. This width can be compared to those of the IR emission lines, such as Fe II or Pa$\beta$. These have been spatially resolved, and have peak-to-peak amplitude velocity differences of $\approx 200~\mathrm{km}~\mathrm{s}^{-1}$ \citep{2001ApJ...549..915M}. However, within narrow slits, values up to $\mathrm{FWHM} \approx 600~\mathrm{km}~\mathrm{s}^{-1}$ were also found \citep{2001ApJ...549..915M}. Additionally, emission at different radii can produce similar FWHMs for different inclinations. Therefore, the narrow cores of the Fe K$\alpha$ line may be produced at comparable, or smaller radii than the IR emission lines. 

The line consists of components emitted from a large range of different radii from the black hole. The emitting region might be associated with a possible BLR, a torus, and the molecular disk. The narrow cores of the Fe K$\alpha$ complex have comparable widths to the IR and optical emission lines. Therefore, they might be emitted by the same molecular disk. High-resolution X-ray images observed by Chandra cannot resolve a distinct Fe-line emitting region. This is consistent with the outer radii fitted by the model, which are comparable to the angular resolution of Chandra. 

The broad line can only be explained by the existence of high-velocity gas, such as exists in a BLR, or something comparable to it. However, \citet{2025arXiv250205002M} found that optical spectropolarimetric data from Cen A did not reveal any indication of a BLR, and that even the presence of a hidden BLR could be excluded with a probability of $\ge99\%$. The physical origin of the broad Fe K$\alpha$ feature is currently unclear. 

Although there is significant uncertainty on the optical depth of the line-emitting region, all spectral fits indicate that it is Compton-thin. The Compton shoulder appears to be weak, and could not be resolved from the broad line that extends both towards higher and lower energies. However, the existence of the broad line may complicate the modeling of the Compton shoulder. 

When assuming a consistent density, the line is well-fitted with an emissivity index of $q\approx2$. This is true even when allowing for some distinct emission regions. Fits to the line with other emissivity indices require non-uniform densities. For instance, when setting $q=3$, we find that three different emission regions are required to describe the line profile with an accuracy comparable to the fit with a single uniform emission region, with $q=2$.

Models A, B, and C all find $q\approx2$. This emissivity index requires either an extended corona, a torus that bends upwards to maintain a fixed angle to the corona, a density that increases with radius, or a combination of these factors. The constraint on the emissivity index comes predominantly from the shape of the broad line. Outer line-emitting regions that produce the narrow peaks might have a different emissivity index, but this cannot be accurately constrained. The emissivity index may be biased by the choice of spectral models used. However, for the measured equatorial $N_{\rm H}$, we expect most Fe K$\alpha$ photons to escape the reprocessor before interacting. A more accurate physical model would find the incident photon flux to decrease faster with radius than assumed by the models we used. This is due to more incident photons being absorbed or scattered with increasing radius. Therefore, an even smaller value of $q$ would be needed to compensate for this effect, and still fit the same observed line shape.

The large relative normalization parameter required by model A is offset in more complex models through the inclusion of additional line-emitting regions. Nevertheless, due to constraints on the radii of the individual components, two line components do not fit the line in the same way that a single line component with twice the relative normalization constant would fit it.

The line can be fitted with two narrow rings at radii equivalent to those found from reverberation mapping, if a third inner, and extended emission region is included. However, it should be noted that there is a lot of freedom when fitting this model. Two narrow rings at different radii and a third extended component can fit almost any line shape well. The uncertainty in the black hole mass, coupled with the uncertainty in the measured reverberation mapping radii, do not provide very tight constraints. Furthermore, there are disagreements about the strengths of the different line-emitting components between the results from reverberation mapping and the requirements from line fitting. This may, however, be partially due to long-term variability, and the chance detection of comparatively weaker emission from a $\approx0.19~\mathrm{pc}$ radius ring. Future modeling of the Fe K$\alpha$ line in the context of other AGN emission lines may provide better constraints on its origin. 

Regardless of the model used, we found that the Fe K$\alpha$ line is redshifted by $(1.89\pm0.03)\times10^{-3}$. This is slightly more than the redshift of Cen A, which is $(1.819\pm0.010)\times10^{-3}$. The origin of this difference of $(7\pm3)\times10^{-5}$, or $20\pm7 ~\mathrm{km}~\mathrm{s}^{-1}$  may be partly due to spacecraft motion in its Earth orbit, but is consistent with systematic gain errors.  This difference is significantly smaller than that measured by \citet{2024ApJ...961..150B}.

We investigated to what extent the different measurements of the redshift could be explained by the different resolutions of Chandra/HETG and XRISM/Resolve. However, when binning the XRISM/Resolve spectrum to match the Chandra resolution, and fitting with the same models as \citet{2024ApJ...961..150B}, we found the same redshift as was measured for the non-rebinned XRISM/Resolve spectrum. It is unclear what caused this difference in measured redshift. 

\citet{2024ApJ...973L..25X} previously found that the Fe K$\alpha$ line observed by XRISM/Resolve in NGC 4151 could only be well-fitted by including three \texttt{MYTorusL} lines with $q=3$ in the model. They associated these three components with the disk, the torus, and the broad-line region. This is similar to the result of this investigation, which found that the Cen A Fe K$\alpha$ line also requires a narrow, an intermediate, and a broad line element to fit, when setting $q=3$. Using the same three-component model, but setting $q=2$, \citet{2024ApJ...973L..25X} found consistent best-fit parameters, except for the inclination. In contrast, we found that when the emissivity index was left unconstrained, the line shape could be adequately described with just a single line component and $q=1.99\pm0.03$. Whereas \citet{2024ApJ...973L..25X} found inconsistent inclinations for the three distinct line-emitting components, we instead found consistent values for the two or three components. There are further similarities between the best-fit properties in NGC 4151 and Cen A, including a low best-fit inclination, as well as a Compton-thin $N_{\rm H}$ of the Fe line emitting structures.

The Hydrogen column densities of models B, C, and D are unreliable, as they either account for the drop in emissivity with radius, or were used as a proxy for both the $N_{\rm H}$, and the relative normalization of each of the \texttt{MYTorusL} components. The Hydrogen column density needed to reproduce the shape and flux of the line with model A is $10^{+6}_{-10}\times10^{22}~\mathrm{cm}^{-2}$. This roughly agrees with the line-of-sight (LOS) $N_{\rm H}$ that is typically measured for Cen A \citep{2011ApJ...733...23R, 2024ApJ...961..150B}. However, the absorption of the power-law spectrum is likely to be both due to material in the Fe line-emitting regions, as well as from the molecular dust lane at radii of tens of parsecs \citep{2017ApJ...843..136E, 2017ApJ...851...76M}. The best-fit low inclination of the line-emitting region would indicate that it only contributes a small amount to the LOS absorption. The relationship between the column densities required by the spectral lines, and those that absorb the power-law continuum will be investigated further when considering the broad-band spectrum observed by XRISM, NuSTAR, and Chandra. 

All of the models we used were best fit with a low inclination of $\approx23-30\degree$, and with $1\sigma$ errors spanning the range of $17\degree-42\degree$. This result agrees well with some of the measured inclination of the jet on small \citep{2014A&A...569A.115M, 2021NatAs...5.1017J} and large \citep{2003ApJ...593..169H, 2024ApJ...974..307B} scales. This also agrees with some of the measured inclinations of the inner molecular disk on parsec scales \citep{2007MNRAS.374..385K, 2007ApJ...671.1329N, 2010PASA...27..396Q}. However, it disagrees with the optical classification of Cen A as a Seyfert-2 galaxy, and also disagrees with the inclination of the outer molecular disk \citep{2009ApJ...695..116E, 2009ApJ...702.1127R}. 

For models A and B, the inclination constraint comes predominantly from fitting the broad line. When the inclination of the components producing the narrow core is allowed to vary freely, it is not well constrained. In models A and B, an inclination of $90\degree$ for all components is only excluded at a $\approx 1.6\sigma$ level. In model C, however, the inclination constraint also comes from the reverberation mapping radii. For this model, an inclination of all components of $90\degree$ is ruled out with a $6.7\sigma$ significance. Model D also rules out a high inclination, due to the constraint on the emissivity, and the requirements of each component to fit distinct parts of the line profile. 

\section{Conclusions}\label{sec:conc}

The key findings of this paper can be summarized as: 

\begin{itemize}
    \item The Cen A Fe K$\alpha$ line observed by XRISM/Resolve consists of two narrow peaks for Fe K$\alpha_1$ and Fe K$\alpha_2$, as well as a broad line.  The narrow peaks have FWHM~$=(4.8\pm0.2)\times10^2~\mathrm{km}~\mathrm{s}^{-1}$. The broad component has a FWHM velocity of $ (4.3\pm0.3)\times10^3~\mathrm{km}~\mathrm{s}^{-1}$, which does not have a corresponding component in the IR or optical spectrum. The ratio of the flux of the narrow lines to the flux of the broad line is $1.07\pm0.09$, with a total flux of $(3.2\pm0.2)\times10^{-12}~\mathrm{erg}~\mathrm{cm}^{-2}~ \mathrm{s}^{-1}$, and an equivalent width of $96\pm4~\mathrm{eV}$.  
    \item The best fit is obtained by fitting the line profile with a single \texttt{MYTorusL} line produced by an extended range of radii ranging from $5.4^{+7.8}_{-4.0}\times10^2~r_{\rm g} = 1.4^{+2.1}_{-1.1}\times10^{-3}~\mathrm{pc}$ to $6.6^{+13.8}_{-3.2}\times10^6~r_{\rm g} = 17^{+36}_{-9}~\mathrm{pc}$. The emissivity decreases with radius with a power-law index of $q=1.99\pm0.03$, and the inclination is $24^{+13}_{-7}~\degree$. The best fit with this model requires a large relative normalization of the line, but this can be accounted for by past variability, lags between the line and the continuum, and possibly different abundances, opening angles, enhanced dust, or different geometries. 
    \item Other, more complex models were fit with comparable inner and outer radii, emissivity indices, and inclinations. They can account for differences in the uniform extended toroidal structure. 
    \item A model based on emission from specific radii measured by reverberation mapping can also yield an accurate description of the line. However, this requires the inclusion of a third emission region that is closer to the black hole.
    \item Alternatively, when setting the emissivity index to $q=3$, the line profile requires three separate line-emitting components to be fitted accurately. The best fit with this model compensates for the higher emissivity index by increasing the density with radius. 
    \item The range of radii required to fit the Fe K$\alpha$ line profile indicates an extended emission that may be associated with a high velocity structure, such as a BLR seen in Seyfert I galaxies and a molecular dust lane. 
    \item We measured a model-independent low inclination of $\approx 25\degree$, consistent with the inclination of the jet, and the inner molecular disk that produces IR lines such as Pa$\beta$ or Fe II. However, a low inclination is only required to fit the broad line. The narrow cores could instead be produced at a higher inclination.
    \item The simplest fits of the line are obtained with an emissivity index of $q\approx2$. This emissivity index requires either an extended corona, an emitting region that bends upwards with radius, or a density that increases with radius.
    \item  Production of the Fe K$\alpha$ line requires a Compton-thin emission region consistent with the weak Compton shoulder in the data. 
\end{itemize}

\section{Acknowledgements}

\noindent
This research has made use of data obtained from the Chandra Data Archive and the Chandra Source Catalog, and software provided by the Chandra X-ray Center (CXC) in the application packages CIAO and Sherpa.

This work made use of the software packages \texttt{astropy} (\url{https://www.astropy.org/}; \cite{2013A&A...558A..33A, 2018AJ....156..123A, 2022ApJ...935..167A}), \texttt{numpy} (\url{https://www.numpy.org/}; \cite{harris2020array}), \texttt{matplotlib} (\url{https://matplotlib.org/}; \cite{Hunter:2007}), and \texttt{scipy} (\url{https://scipy.org/}; \cite{2020SciPy-NMeth}).

TY acknowledges support by NASA under award number 80GSFC24M0006. LCG acknowledges financial support from the Canadian Space Agency grant 18XARMSTMA. This work was supported by JST SPRING, Grant Number JPMJSP2132. HN is supported by Japan Society for the Promotion of Science (JSPS) KAKENHI with Grant numbers of 19K21884, 20H01941, 20H01947, 20KK0071, 23K20239, and 24K00672. This work was supported by JSPS KAKENHI Grant Number JP23KJ0780 (TI). This work was supported by the Grant-in-Aid for JSPS Fellows (YN). This work was supported by the Grant-in-Aid for Scientific Research 20H01946 (YU). This work was supported by JSPS KAKENHI Grant Number JP21K13958 (MM) and Yamada Science Foundation (MM). The material is based upon work supported by NASA under award number 80GSFC24M0006.

We thank Yuichi Terashima for his helpful comments. 

\bibliography{bibliography}
\bibliographystyle{aasjournal}

\appendix

\section{Investigating a possible additional line component}\label{sec:4MYT}

As Figs. \ref{fig:FeKline_2MYT}, \ref{fig:FeKline_Iwata}, and \ref{fig:FeKline_3MYT} show, there are some further discrepancies between the data and the model at the lowest and highest energies of the broad tails of the Fe K$\alpha$ line profile. We found that some of these features can simultaneously be fitted by adding one additional blurred \texttt{MYTorusL} line to the model that is located at small radii, and has a narrow radial extent. 

Here we will present the results of adding this additional component to model A. It can also be added to the other models as well, with a similar impact on the fits. 

The additional component has a low flux, so there is an even larger degeneracy between the relative normalization and the Hydrogen column density than for the two components in model B. Therefore, we set the multiplicative constant of this extra component equal to 1. Besides that, the full model we investigated here has the same XSPEC notation as model B. 

\begin{figure}[t]
\begin{center}
\resizebox{\hsize}{!}{\includegraphics{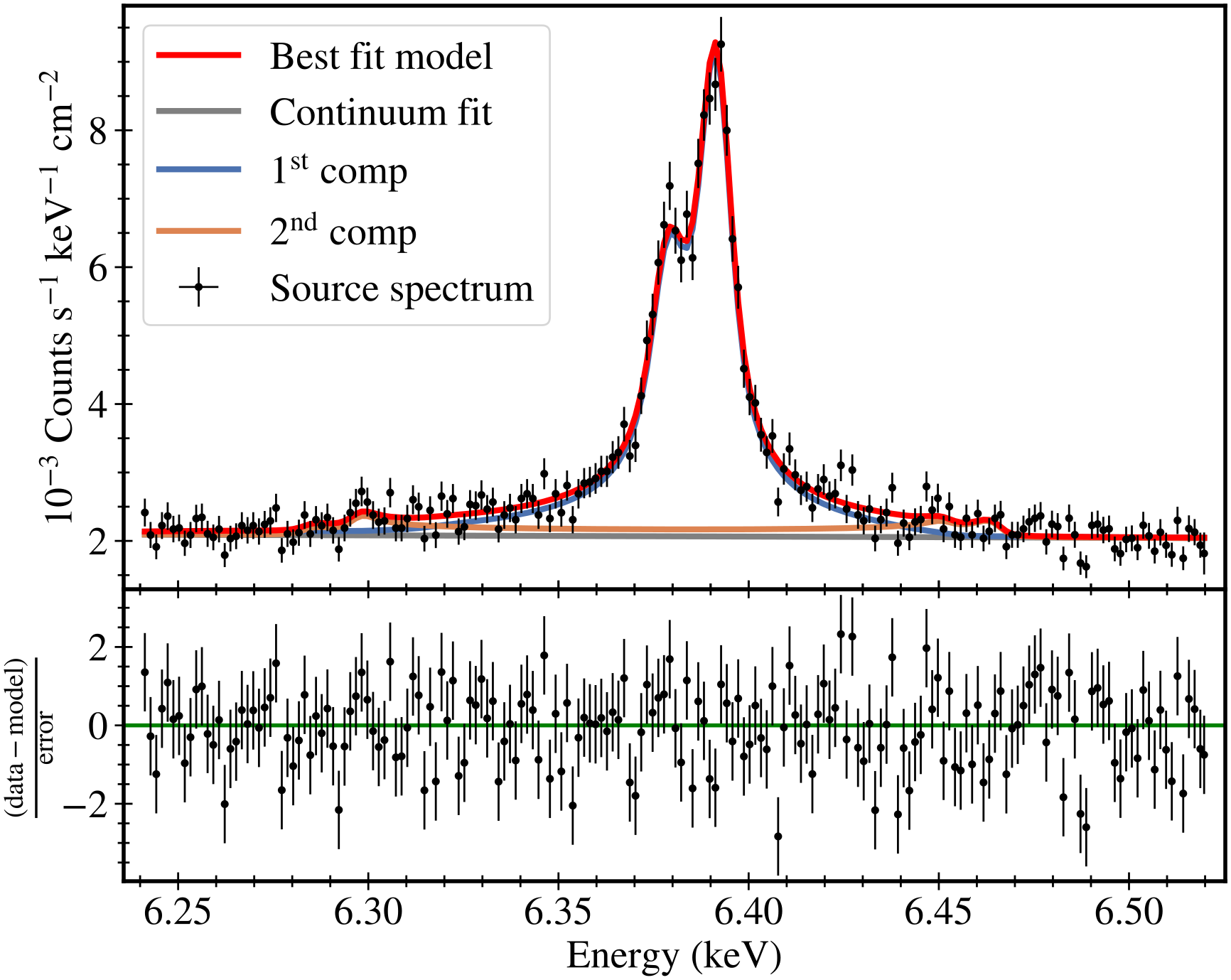}}
\end{center}
\caption{XRISM/Resolve spectrum of the Fe K$\alpha$ line, fitted using model B, but with the second component having a very small radial extent, and only fitting the broadest parts of the line profile.
{Alt text: Figure of the fitted spectrum, and the ratio of the difference between data and model, and the error of each data point. The figure shows the best fit with this model, and also depicts the two blurred line components that contribute to it. The continuum below the line is also shown.}  
 \label{fig:FeKline_1MYTring}}
\end{figure}

Fig. \ref{fig:FeKline_1MYTring} displays the best fit found with this model. The additional \texttt{MYTorusL} component can simultaneously fit some of the largest deviations between the best fit of model A and the line profile, at both the high and low energy ends of the line shape. The resulting C-statistic is $1656.76$, with $1584$ degrees of freedom, and a corresponding $\Delta\mathrm{BIC} = 14.17$, relative to the best-fit with model A.

This is a lower C-statistic than that of any other model presented here, and has a BIC lower than the best-fit with models B, C, and D. However, the BIC is kept artificially low through the exclusion of the relative normalization parameter for this second component. It is also not statistically preferred over model A, as it requires 3 additional fitting parameters.

However, the best-fit parameters of the second line-component are challenging to physically account for. The inner radius is $8.2\times10^2~r_{\rm g}$, but its outer radius is fitted to be virtually identical, also at $8.2\times10^2~r_{\rm g}$. This component is fitted with a Hydrogen column density of $\approx5\times10^{22}~\mathrm{cm}^{-2}$. This would correspond to a significantly larger volume density for this ring-like structure with a limited radial extent, compared to the main extended emission region. This could potentially be explained as a tidally disrupted structure in a circular orbit. It has an equivalent width of $15^{+10}_{-5}~\mathrm{eV}$.

Most of the best-fit parameters of the main line profile with this model are consistent with those found when fitting with model A. The inner radius of the main outer component is now found at a larger, but still consistent radius, of $\approx1.2\times10^3~r_{\rm g}$. The inner line component requires a low inclination. For the case of identical inclinations for both components, the best fit finds $i\approx22\degree$.

We also investigated whether the initial assumption of identical inclination of the two components was valid for this model. When allowing the inclinations to vary freely, we found that the outer component was best fit with $i\approx 85\degree$, and the inner component with $i\approx10\degree$. In this fit, the inner ring fitted other parts of the high-energy tail, and the C-statistic was further reduced by $2.11$. Neither inclination is well-constrained. Nevertheless, this indicates that most of the Fe K$\alpha$ line-emitting region can have a high inclination, as long as there is also a low-inclination line-emitting structure at small radii. The low inclination found by models A, B, and C is predominantly due to the requirements needed to fit the broad wings of the line. 

We also considered the possibility of multiple ring-like structures akin to the one described here. However, the inclusion of another model like this did not result in any significant improvement to the fit statistic. 

It is unclear if this inner narrow-ring-like structure exists, and its required physical parameters are questionable. Additional data would be needed to better constrain the geometry of the inner components that produce the Fe K$\alpha$ line, and look for any inhomogeneities like this one. Nevertheless, the ability to investigate models like this showcases the unprecedented sensitivity achieved by XRISM/Resolve, and the new ability to investigate and identify structures that were previously undetectable. 

\end{document}